\date{\today}

\documentclass[aps,pra,twocolumn,superscriptaddress,groupedaddress]{revtex4}  % for review and submission
\usepackage{graphicx,amsmath,amssymb,algorithm,algpseudocode}
\usepackage{todonotes}
\usepackage{qcircuit}
\usepackage{mathtools}
\usepackage[colorlinks=true,hyperindex,breaklinks=true,allcolors=blue]{hyperref}

\newcommand{\ud}{\mathrm{d}}

\newcommand{\erfc}{\mathrm{erfc}}
\newcommand{\diff}[2]{\frac{\ud {#1}}{\ud {#2}}}
\newcommand{\pdiff}[2]{\frac{\partial #1}{\partial #2}}

\begin{document}

\definecolor{brickred}{rgb}{.72,0,0} 
\definecolor{darkblue}{rgb}{0,0,0.5} 
\definecolor{darkgreen}{rgb}{0,0.5,0} 

\title{
% Molecular ground and excited state analytical gradients by om hybrid quantum classical 

% Analytical Energy Gradients for the Multistate, Contracted Variational Quantum Eigensolver 
% for Fermionic Systems

% First-Order Gradients for Hybrid Quantum/Classical Methods for Molecular
% Electronic Structure Theory

Analytical Ground- and Excited-State Gradients for Molecular Electronic
Structure Theory from Hybrid Quantum/Classical Methods
}

\author{Robert M. Parrish}
\email{rob.parrish@qcware.com}
\affiliation{
QC Ware Corporation, Palo Alto, CA 94301, USA
}

\author{Gian-Luca R. Anselmetti}
\email{gian-luca.anselmetti@covestro.com}
\affiliation{
Covestro Deutschland AG, Leverkusen 51373, Germany
}

\author{Christian Gogolin}
\email{christian.gogolin@covestro.com}
\affiliation{
Covestro Deutschland AG, Leverkusen 51373, Germany
}

\begin{abstract} 
We develop analytical gradients of ground- and excited-state energies with
respect to system parameters including the nuclear coordinates for the hybrid
quantum/classical multistate contracted variational quantum eigensolver (MC-VQE)
applied to fermionic systems. We show how the resulting response contributions
to the gradient can be evaluated with a quantum effort similar to that of obtaining 
the VQE energy and independent of the total number of derivative parameters
(e.g. number of nuclear coordinates) by adopting a Lagrangian formalism for the
evaluation of the total derivative. We also demonstrate that large-step-size
finite-difference treatment of directional derivatives in concert with the parameter
shift rule can significantly mitigate the complexity of dealing with the quantum
parameter Hessian when solving the quantum response equations. This enables the
computation of analytical derivative properties of systems with hundreds of
atoms, while solving an active space of their most strongly correlated orbitals
on a quantum computer. We numerically demonstrate the exactness the analytical
gradients and discuss the magnitude of the quantum response contributions.
\end{abstract}

\maketitle

\section{Introduction}

When performing simulations, computational chemists frequently depend on a
toolset of computational methods to provide observables beyond approximations
to the state energies for a given arrangement of nuclear cores and electronic
quantum numbers.  First-order derivatives of the energy and other properties of
ground- and excited-state electronic wave functions with respect to various
system parameters are frequently sought after quantities. Of particularly broad
interest is the nuclear gradient, i.e., the first order derivative of the state
energy with respect to the nuclear positions. This nuclear gradient represents
the classical force acting on point Born-Oppenheimer nuclei, and is
indispensable for the determination of critical points on the potential energy
surface, the exploration of reaction landscapes, and the computation of static
and dynamic properties of molecular systems.

Due to the necessary layering of different numerical techniques, e.g., the
combination of a Hartree-Fock-type method to identify molecular orbitals combined
with a post-Hartree-Fock-type method to provide an accurate treatment of the
electronic wavefunctions, the usage of intermediate computational parameters is
necessary. The values of these parameters often need to be determined by means of
nested optimization loops which solve particular sets of nonlinear equations.
This makes precise computation of derivatives challenging in both existing
classical and forthcoming quantum methods, as ``response'' cross terms stemming
from the chain rule derivatives of the intermediate computational parameters
will necessarily contribute to the target total derivatives. In the quantum
methods, the difficulty is compounded due to the inability to directly access
the wave function.

In classical methods, the efficient computation of the full analytical
derivative of an observable with respect to arbitrary system parameters, and
particularly including response contributions from the chain-rule terms
involving the intermediate computational parameters, has been thoroughly
established over the past several decades. A particularly efficient and robust
approach for deriving and implementing these gradients is present in the
``Lagrangian formalism'' promoted by Helgaker and co-workers
\cite{Helgaker:1982:LAG}, which itself grew out of several other historical
perspectives on the topic, including the Delgarno-Stewart interchange theorem
\cite{Dalgarno:1958:245}, the Handy-Schaefer Z-vector formalism
\cite{Handy:1984:Zvec}, and the
coupled-perturbed approach used by Yamaguchi and many others
\cite{Schaefer:1986:new}. It remains, however, to fully extend these efficient
classical techniques for analytical gradients into the domain of hybrid
quantum/classical algorithms for quantum chemistry.

A number of notable studies have already made considerable progress in computing
derivatives in hybrid quantum/classical methods for quantum chemistry. In
quantum phase estimation (QPE) methods, approaches for the derivative have been
considered more than a decade ago \cite{Kassal:2009:grad}. Our manuscript
focuses primarily on methods for NISQ-era \cite{preskill2018quantum} quantum
computing, in which many different variational, semi-variational, or
non-variational methods for the state energy have been developed
\cite{
mcclean2017hybrid,
parrish2019quantum,
nakanishi2019subspace,
parrish2019qfd,
urbanek2020chemistry,
ollitrault2020quantum,
huggins2020non,
stair2020multireference,
klymko2021real}.
In these cases, depending on the variational nature of the method with regard
to the parameter(s) to be differentiated, considerable additional response
terms may arise in the computation of the analytical gradient.  the earliest
days of the variational quantum eigensolver (VQE) approach, it has been known
to many authors that if only the ground state energy is targeted, the nuclear
gradient is analytically extractable from the active space density matrix
elements that are used already in the optimization of the VQE entangler
parameters, i.e., that the gradient is essentially available simultaneously
with the energy due to the variational nature of state-specific VQE
\cite{McClean:2016:023023}.  However, when one considers approaches where the
VQE entangler parameters are not optimized to minimize the observable to be
differentiated, the resulting method will be non-variational, and nonzero chain
rule terms will arise to reflect the response of the VQE entangler parameters
to the derivative perturbations. This case commonly arises in methods that
target ground and excited states simultaneously, e.g., MC-VQE
\cite{parrish2019quantum} or SS-VQE \cite{nakanishi2019subspace} (in which case the
state-averaged energy is optimized by tuning the VQE entangler parameters in an
approach denoted SA-VQE), or in cases where an observable other than the energy
is differentiated (e.g., in the computation of polarizabilities or
non-adiabatic coupling vectors in state-specific VQE). There may also be
response terms stemming from parameters other than the VQE entangler
parameters, such as orbital parameters.

There seem
to be three broad approaches to dealing with the response terms. The first is to
simply ignore the response contribution from to the VQE entangler parameters, noting that in the limit of a universal entangler circuit, the
response terms will decay to zero, and that there will be competing error
sources from shot and decoherence noise that may overwhelm the magnitude of the
response terms. The second is to redefine the method to eliminate the response
terms. This was done, e.g., in \cite{delgado2021variational}, where a state-specific CASSCF was performed to
simultaneously optimize the state-specific VQE energy with respect to both the
VQE entangler parameters and the orbital parameters. This yields a method which
is fully variational, and which in particular does not require the solution of
the classical coupled-perturbed Hartree-Fock (CPHF) equations to determine the
orbital response terms. However, this method will encounter problems if one
tries to differentiate observables other than the state-specific energy or if
state-specific CASSCF does not provide a reliable treatment of the electronic
structure.The third is to consider the response terms as a natural consequence
of a non-variational method, and to try to efficiently include their
contribution in the computation of the derivative. Varying levels of efficiency
are obtained depending on the formalism used. For instance, a pair of studies
of analytical gradients in the non-variational SS-VQE method
\cite{mitarai2020theory,Azad2021} explicitly compute the response terms,
achieving a correct analytical gradient but with quantum computational cost
scaling linearly in the total number of derivative directions. Another study
focusing on the non-variational QSE-VQE approach \cite{o2019calculating}
includes the response contributions through a sum-over-states formalism that
also induces significant overhead. Recently, we have begun the exploration of
the Lagrangian approach for analytical derivatives in the context of the
phenomonological \emph{ab initio} exciton model (AIEM) solved with MC-VQE
\cite{parrish2019hybrid}. There we showed how the Lagrangian formalism could
sever the dependency of the computational scaling of the gradient from the
number of derivative parameters being sought. However, we still encountered
significant computational complexity increase from the nature of the SA-VQE
Hessian matrix, which is seemingly much harder to deal with in quantum methods
relative to classical methods. We make some progress to mitigate this issue in
the present manuscript.

As we were finalizing the present manuscript, a preprint by Arimitsu \emph{et
al} appeared \cite{arimitsu2021analytic}, which derives and implements the
gradients of orbital-optimized fermionic SS-VQE and MC-VQE through a Lagrangian
formalism. Similar response equations  are encountered as in our prior
derivative work \cite{parrish2019hybrid} or as derived below, but the authors
consider neither previous approaches introducing the Lagrangian formalism for
non-variational gradients in hybrid quantum/classical methods, nor efficient
matrix-vector product techniques for mitigating the complexity of working with
the SA-VQE parameter Hessian. The latter consideration, together with
unavoidable couplings between the quantum VQE parameters and classical orbital
parameters manifests as a restriction of the Arimitsu work to small test cases
such as 9-atom cis-TFP. A similar Lagrangian-based approach for the
non-adiabatic coupling vector has since appeared \cite{yalouz2021analytical}.

In the present manuscript, we explore the deployment of the Lagrangian approach
to provide the analytical gradient for non-variational hybrid quantum/classical
methods with fermionic Hamiltonians. We particularly consider the case where
there are nonzero response terms arising from both classical intermediate
systems parameters such as orbitals determined by Hartree-Fock-type methods, as
well as quantum intermediate system parameters such as VQE entangler circuit
parameters determined by SA-VQE. For concreteness, we demonstrate numerical
results for the case where the orbitals are chosen by the fractional occupation
molecular orbital restricted Hartree-Fock (FON-RHF) method \cite{gidopoulos1994hartree,gidopoulos2002ensemble}, and then the
ground and excited state wavefunctions are simultaneously optimized in an
active space picture via the multistate, contracted variant of VQE (MC-VQE)
This method is an analog to the popular FOMO-CASCI method \cite{barbatti2006ultrafast,sellner2013ultrafast} of classical
electronic structure, for which analytical gradients and non-adiabatic coupling
vectors have recently been derived \cite{hohenstein2015analytic,hohenstein2016analytic}.
However, we leave sufficient detail in the theory section to show how the same
approach could be adapted to many other combinations of classical and quantum
methods, especially including the case where the quantum and classical
intermediate system parameters are globally coupled. We also find that our
recently developed quantum number preserving (QNP) approach for efficient
fermionic VQE entanglers \cite{QNP2021} (itself inspired by and derived from a number of other fermionic VQE entangler approaches 
\cite{peruzzo2014variational,
PhysRevX.6.031007,
Kandala2017,
Ryabinkin2018,
lee2018generalized,
evangelista2019exact,
ganzhorn2019gate,
salis2019short,
Bian2019,
o2019generalized,
gard2020efficient,
xia2020qubit,
yordanov2020efficient,
khamoshi2020correlating,
matsuzawa2020jastrow}) 
fits particularly nicely into this
framework and allow to obtain high quality ground and excited state properties
already at low circuit depth, while also obviating the need to consider any
constraints to keep the system responses from falling into symmetry-impure
sectors of the qubit Hilbert space.

We also make progress with the SA-VQE Hessian complexity problem encountered in
the AIEM gradient manuscript, by showing that a combination of iterative
matrix-vector-product-based solution of the SA-VQE response equations with a
widely spaced finite difference approach for the SA-VQE matrix-vector products
works well in numerical practice. We close by numerically demonstrating the
exactness of our method against  finite difference, showing the performance of the
iterative finite-difference Hessian-vector product method, and exploring the
magnitude of the response terms in large test cases.

\section{Theory}

\subsection{Lagrangian Formalism for Derivative Properties}

As with all gradient theory papers, the topic of the present manuscript is how
to efficiently compute derivatives of observables of a quantum system with
respect to system parameters, while analytically accounting for any non-physical
choices made during the approximate computational description of the quantum
system. To simplify the discussion, we will confine ourselves to the common case
where the observable in question is the adiabatic (ground or excited) state
energy $E^{\Theta}$ corresponding to an orthonormal adiabatic state $|
\Psi^{\Theta} \rangle$. We will also restrict ourselves to the symmetric and
linear ansatz with the property that the adiabatic energy is the expectation
value over the Hamiltonian operator $\hat H$
\footnote{This covers most
present hybrid quantum/classical approaches deployed within active spaces,
including MC-VQE and related VQE variants, as well as QFD and related VQPE
variants, but does not cover nonsymmetric methods like (classical) projective
coupled cluster theory.}.
\begin{equation}
E^{\Theta}
\equiv
\langle \Psi^{\Theta} |
\hat H
| \Psi^{\Theta} \rangle
\end{equation}
We often require total derivatives of this adiabatic state energy with respect to
arbitrary displacements of system parameters of the Hamiltonian $\{ \xi \}$.
An archetypical example of $\{ \xi \}$ is the set of Cartesian positions of the
positively-charged nuclei of a molecular system, in which case $E^{\Theta, \xi}$
represents the nuclear gradient of the energy, i.e., the opposite of the
classical Born-Oppenheimer force on the nuclei. The desired derivative is,
\begin{equation}
E^{\Theta, \xi}
\equiv
\diff{E^{\Theta}}{\xi}
=
\langle \Psi^{\Theta} |
\pdiff{\hat H}{\xi}
| \Psi^{\Theta} \rangle
+
\langle \Psi^{\Theta} |
\hat H
| \diff{\Psi^{\Theta}}{\xi} \rangle
+
\mathrm{H.C.}
\end{equation}
Here, the first term is the Hellmann-Feynman contribution, and is the only
nonzero contribution if all parameters in $| \Psi^{\Theta} \rangle$ were chosen
to make the state energy $E^{\Theta}$ stationary (definitionally true for all
state-specific variational wavefunctions). However, if the wave function contains
intermediate computational parameters $\{ \theta_{g} \}$ that were chosen to
satisfy other conditions than the stationarity of the state energy, the
``wave function response'' terms will arise,
\begin{equation}
\begin{split}
E^{\Theta, \xi}
\equiv
\diff{E^{\Theta}}{\xi}
=
\langle \Psi^{\Theta} |
\pdiff{\hat H}{\xi}
| \Psi^{\Theta} \rangle
\\
+
\sum_{g}
\langle \Psi^{\Theta} |
\hat H
| \pdiff{\Psi^{\Theta} (\{ \theta_g \})}{\theta_g} \rangle
\pdiff{\theta_{g}}{\xi}
+
\mathrm{H.C.}
\end{split}
\end{equation}
Structurally identical terms arise whenever the quantity of which the derivative 
is taken is not the one whose optimization yielded the intermediate computational
parameters $\{\theta_{g}\}$.

At the moment, it seems that we will have to solve the the parameter response
$\pdiff{\theta_{g}}{\xi}$ separately for each perturbation $\xi$, which will
induce unacceptable rise in complexity of the gradient computation relative to
the energy computation. Fortunately, the Lagrangian formalism helps us avoid
this. We define,
\begin{equation}
\mathcal{L}^{\Theta}
\equiv
\langle \Psi^{\Theta} |
\hat H
| \Psi^{\Theta} \rangle
+
\underbrace{
\sum_{g'}
\tilde \theta_{g'}
C_{g'} (\{ \theta_{g} \} )
}_{0}
\end{equation}
Here $C_{g'} (\{ \theta_{g} \}) = 0$ is the $g'$-th clause of a set of $n_{g}$
(nonlinear) equations used to define the parameters $\{ \theta_{g} \}$. $\{
\tilde \theta_{g} \}$ are Lagrange multipliers. Making the Lagrangian stationary
with respect to the Lagrange multipliers specifies the method,
\begin{equation}
\pdiff{\mathcal{L}^{\Theta}}{\tilde \theta_{g'}}
=
0
\Rightarrow
C_{g'} (\{ \theta_{g} \}) 
=
0
\end{equation}
Making the Lagrangian stationary with respect to the wave function parameters
determines the linear response equations,
\begin{equation}
\pdiff{\mathcal{L}^{\Theta}}{\theta_{g}}
=
0
\Rightarrow
\sum_{g'}
\pdiff{C_{g'} (\{ \theta_{g} \})}{\theta_{g}}
\tilde \theta_{g'}
=
-
\pdiff{
\langle \Psi^{\Theta} |
\hat H
| \Psi^{\Theta} \rangle
}{\theta_{g}}
\end{equation}
Once the Lagrangian (whose value always equals the energy) is stationary with
respect to perturbations in non-variational wave function parameters, the
desired derivatives can be taken as partial rather than total derivatives,
\begin{align}
E^{\Theta, \xi}
&=
\diff{E^{\Theta}}{\xi}
=
\pdiff{\mathcal{L}}{\hat H}
\diff{\hat H}{\xi}\\
&=
\langle \Psi^{\Theta} |
\pdiff{\hat H}{\xi}
| \Psi^{\Theta} \rangle
+
\sum_{g'}
\tilde \theta_{g'}
\pdiff{C_{g'} (\{ \theta_{g} \})}{\hat H}
\pdiff{\hat H}{\xi}\\
&=
\left (
\pdiff{E^{\Theta}}{\hat H}
+
\sum_{g'}
\tilde \theta_{g'}
\pdiff{C_{g'} (\{ \theta_{g} \})}{\hat H}
\right)
\pdiff{\hat H}{\xi}
\end{align}
The quantities $\pdiff{E^{\Theta}}{\hat H}$ and $\pdiff{\mathcal{L}^\Theta}{\hat
H} = \diff{E^{\Theta}}{\hat H}$ are called the
``unrelaxed'' and ``relaxed'' density matrices.

Note that the notation $\pdiff{}{\hat H}$ is meant to imply a sum over the
partials of the linear matrix element coefficients in $\hat H$.

\subsubsection{Nested Classical/Quantum Ansatze}

It is often the case in hybrid quantum/classical methods (or even
classical/classical methods) that several sequential computations are
concatenated to form a complete method. For one instance, one may perform
some classical computations to determine the spatial orbitals (defined in terms of
classical intermediate computational parameters $\{
\theta_{g}^{\mathrm{c}} \}$), and subsequently use these orbitals in a VQE-type
computation with state-averaged variational entangler circuits (depending on
quantum intermediate computational parameters $\{
\theta_{g}^{\mathrm{q}} \}$). The resultant adiabatic energies $E^{\Theta}$ will
be variational with respect to neither classical nor quantum intermediate
computational parameters. However, the Lagrangian formalism will naturally guide
us to a set of response equations which are succinct and maximally separated.
\begin{equation}
\begin{split}
\mathcal{L}^{\Theta}
\equiv
E^{\Theta} ( 
\{ \theta_{g}^{\mathrm{q}} \},
\{ \theta_{g}^{\mathrm{c}} \}
)
&+
\sum_{g}
\tilde \theta_{g}^{\mathrm{q}}
C_{g}^{\mathrm{q}} (\{ \theta_{g'}^{\mathrm{c}} \}; \{ \theta_{g'}^{\mathrm{c}} \})\\
&+
\sum_{g}
\tilde \theta_{g}^{\mathrm{c}}
C_{g}^{\mathrm{c}} (\{ \theta_{g'}^{\mathrm{c}} \})
\end{split}
\end{equation}
The second line indicates that the equations defining the quantum intermediate
computational parameters depend on both the quantum and classical intermediate
computational parameters, but the equations defining the classical intermediate
computational parameters do not depend on the quantum intermediate computational
parameters. This leads to a nested and separated set of response equations,
\begin{align}
\pdiff{\mathcal{L}^{\Theta}}{\theta_{g}^{\mathrm{q}}}
&=
0
\Rightarrow
\sum_{g'}
\tilde \theta_{g'}^{\mathrm{q}}
\pdiff{
C_{g'}^{\mathrm{q}} 
}{\theta_{g}^{\mathrm{q}}}
=
-
\pdiff{E^{\Theta}}{\theta_{g}^{\mathrm{q}}}
\intertext{and}
\pdiff{\mathcal{L}^{\Theta}}{\theta_{g}^{\mathrm{c}}}
&=
0
\Rightarrow
\sum_{g'}
\tilde \theta_{g'}^{\mathrm{c}}
\pdiff{
C_{g'}^{\mathrm{c}} 
}{\theta_{g}^{\mathrm{c}}}
=
-
\pdiff{E^{\Theta}}{\theta_{g}^{\mathrm{q}}}
-
\sum_{g'}
\tilde \theta_{g'}^{\mathrm{q}}
\pdiff{
C_{g'}^{\mathrm{q}} 
}{\theta_{g}^{\mathrm{c}}} .
\end{align}
Critically, this separates the response computations into quantum and classical
portions, with iterative classical/quantum computations avoided in the latter.

In practice, the extra right-hand-side contributions in the classical response
equations naturally are accounted for if one uses quantum-relaxed density
matrices as input to the classical response equations, as we do below.

Note that if the determination of the classical intermediate computational
parameters somehow relied on quantum intermediate computational parameters
(e.g., if the orbitals were chosen to minimize the quantum state-averaged energy
instead of a classical heuristic), the Lagrangian formalism would guide us to a
single, united set of response equations resembling CP-SA-CASSCF equations. This
would present no conceptual difficulty, but might require considerable
additional hybrid quantum/classical computations to evaluate.

\subsection{Active Space Picture}

We begin from a classically-determined active space of $M$ orthonormal spatial
orbitals $\{ \phi_{p} (\vec r_1) \}$ (defined to be real in this work). For each
of these spatial orbitals, we compose a pair of spin orbitals $\{ \psi_{p} (\vec
x_1) \equiv \phi_{p} (\vec r_1) \alpha (s_1) \}$ and $\{ \psi_{\bar p} (\vec
x_1) \equiv \phi_{p} (\vec r_1) \beta (s_1) \}$, for a total of $2M$ spin
orbitals.  When referring to spin orbitals, we use a bar to denote $\beta$ (spin down)
orbitals and no bar to denote $\alpha$ (spin up) orbitals. We define a set of
fermionic composition operators $\{ \hat p^{\pm} \}$ and $\{ \hat {\bar p}^{\pm}
\}$ for these spin orbitals obeying the usual fermionic anti-commutation
relations.

We are given the classically-determined matrix elements of the active space
Hamiltonian, including the self energy $E_{\mathrm{ext}}$ of the external (classically-determined)
system, the one-body integrals, including the kinetic energy integrals and the potential
integrals of the external system
\begin{align}
(p | \hat h | q)
&\equiv
(p | - \nabla_1^2 / 2 | q)
+
(p | \hat v_{\mathrm{ext}} | q) ,
\\
&\equiv
\int_{\mathbb{R}^3}
\mathrm{d} \vec r_1 \
\phi_{p} (\vec r_1)
\left [
-
\frac{\nabla_1^2}{2}
+
\hat v_{\mathrm{ext}} (\vec r_1)
\right ]
\phi_{q} (\vec r_1)
\end{align}
and the two-body electron repulsion integrals (ERIs),
\begin{equation}
(pq|rs)
\equiv
\iint_{\mathbb{R}^6}
\mathrm{d} \vec r_1 \
\mathrm{d} \vec r_2 \
\phi_{p} (\vec r_1)
\phi_{q} (\vec r_1)
\frac{1}{r_{12}}
\phi_{r} (\vec r_2)
\phi_{s} (\vec r_2)
\end{equation}
In the present work these are all real quantities due to the real nature of the
spatial orbitals. Specific definitions of these integrals depend on the type of
active space embedding used (e.g., Hartree-Fock external particles) and on the
definitions of the spatial orbitals as shown in the integrals over spatial
orbitals. See Appendices \ref{sec:FON-RHF} and \ref{sec:EMB-RHF} for specific
definitions for the case of FON-RHF spatial orbitals and RHF core orbital
embedding, respectively.

For convenience, we define the modified one-electron integrals,
\begin{equation}
(p | \hat \kappa | q)
\equiv
(p | \hat h | q)
-
\frac{1}{2}
\sum_{r}
(pr|rq)
\end{equation}
The Hamiltonian can now be defined as,
\begin{equation}
\hat H
\equiv
E_{\mathrm{ext}}
+
\sum_{pq}
(p | \hat \kappa | q)
\hat E_{pq}^{+}
+
\frac{1}{2}
\sum_{pqrs}
(pq|rs)
\hat E_{pq}^{+}
\hat E_{rs}^{+}
\end{equation}
where the singlet spin-adapted single substitution operator is,
\begin{equation}
\hat E_{pq}^{+}
\equiv
\hat p^{\dagger}
\hat q
+
\hat {\bar p}^{\dagger}
\hat {\bar q}
\end{equation}
We also define the $\alpha$ number, $\beta$ number, and total spin squared
quantum number operators as, respectively,
\begin{align}
\hat N_{\alpha}
&\equiv
\sum_{p}
\hat p^\dagger
\hat p
\\
\hat N_{\alpha} 
&\equiv
\sum_{p}
\hat {\bar p}^\dagger
\hat {\bar p}
\\
\hat S^2
&\equiv 
\hat S_{-}
\hat S_{+}
+
\hat S_{z}
+
\hat S_{z}^2
\intertext{where}
\hat S_{-}
&\equiv
\sum_{p}
\bar p^{\dagger}
p
,
\quad
\hat S_{+}
\equiv
\sum_{p}
p^{\dagger}
\bar p
\end{align}

We are first tasked with solving the time-independent electronic Schr\"odinger
equation within the active space, subject to strict quantum number constraints
($N_{\alpha}$, $N_{\beta}$, $S$),
\begin{align}
\hat H
| \Psi^{\Theta} \rangle
&=
E^{\Theta}
| \Psi^{\Theta} \rangle
:
\\
\langle \Psi^{\Theta} | \Psi^{\Theta'} \rangle 
&=
\delta_{\Theta \Theta'}
\\
\hat N_{\alpha} | \Psi^{\Theta} \rangle
&=
N_{\alpha} | \Psi^{\Theta} \rangle
\\
\hat N_{\beta} | \Psi^{\Theta} \rangle
&=
N_{\beta} | \Psi^{\Theta} \rangle
\\
\hat S^2 | \Psi^{\Theta} \rangle
&=
S / 2 (S / 2 + 1) | \Psi^{\Theta} \rangle
\end{align}
We will construct the wavefunctions $\{ | \Psi^{\Theta} \rangle \}$ and measure
observable properties thereof with the aid of a qubit-based quantum computer.
There are many approaches for this hybrid quantum/classical solution of the
Schr\"odinger equation. For concreteness, herein we use the multistate,
contracted variant of the variational quantum eigensolver (MC-VQE) for its
seamless simultaneous treatment of ground and excited states.

\subsection{Target Gradient}

We are then asked to evaluate the derivative of the state energy $E^{\Theta}$
with respect to an arbitrary system parameter $\xi$ (e.g., the position of one
of the nuclei, an external electric field, etc),
\begin{equation}
E^{\Theta,\xi}
\equiv
\diff{E^{\Theta}}{\xi}
\end{equation}
A key intermediate in the mixed quantum/classical methodology used here is the
spin-summed active space relaxed density matrix, which has one-particle (OPDM) contribution,
\begin{equation}\label{eq:relaxed_opdm}
\bar
\gamma_{pq}^{\Theta}
\equiv
\diff{E^{\Theta}}{(p | \hat h | q)}
\end{equation}
and two-particle (TPDM) contribution,
\begin{equation}\label{eq:relaxed_tpdm}
\bar
\Gamma_{pqrs}^{\Theta}
\equiv
2
\diff{E^{\Theta}}{(pq|rs)}
\end{equation}
These quantities should be understood to be relaxed up through all quantum
intermediate computational parameters (i.e., to be the d\"oppelgangers of full
configuration interaction density matrices in the active space), but do not
include or have any context for classical intermediate computational parameters
such as orbital definitions.

The reason that these contributions are so important is that the linear,
symmetrical expectation value property of our ansatz allows us to write (up
through the quantum intermediate computational parameters),
\begin{equation}
\mathcal{L}^{\Theta}
=
E_{\mathrm{ext}}
+
\sum_{pq}
\bar \gamma_{pq}^{\Theta}
(p | \hat h | q)
+
\frac{1}{2}
\sum_{pqrs}
\bar \Gamma_{pqrs}^{\Theta}
(pq|rs)
\end{equation}
And therefore, after considering the properties of the Lagrangian, the desired
derivative can be found as,
\begin{equation}
E^{\Theta,\xi}
=
\diff{}{\xi}
E_{\mathrm{ext}}
+
\sum_{pq}
\bar \gamma_{pq}^{\Theta}
\diff{}{\xi}
(p | \hat h | q)
+
\frac{1}{2}
\sum_{pqrs}
\bar \Gamma_{pqrs}^{\Theta}
\diff{}{\xi}
(pq|rs)
\end{equation}
Here the total derivative symbol is used when differentiating the active space
molecular integrals to emphasize that these contain both intrinsic and orbital
response terms (both classical). In practice, this last line represents a call
to a classical CASCI gradient computation, with relaxed active space OPDM/TPDM
provided as input.

Note also that we will encounter another set of intermediates in the form of the
unrelaxed OPDM,
\begin{equation}
\gamma_{pq}^{\Theta}
\equiv
\pdiff{E^{\Theta}}{(p | \hat h | q)}
=
\langle \Psi^{\Theta} | \hat E_{pq}^{+} | \Psi^{\Theta} \rangle
\end{equation}
and TPDM,
\begin{equation}
\Gamma_{pqrs}^{\Theta}
\equiv
2
\pdiff{E^{\Theta}}{(pq|rs)}
=
\langle \Psi^{\Theta} | \hat E_{pq}^{+} \hat E_{rs}^{+} | \Psi^{\Theta} \rangle
\end{equation}
\[
-
\mathcal{S}
\delta_{qr}
\langle \Psi^{\Theta} | \hat E_{ps}^{+} | \Psi^{\Theta} \rangle
\]
Here the symmetrization operator (a convention) acts as,
\begin{equation}
\mathcal{S} T_{pqrs}
=
\frac{1}{4}
\left [
T_{pqrs}
+
T_{pqsr}
+
T_{qprs}
+
T_{qpsr}
\right ]
\end{equation}
These do not include any relaxation terms within the active space, e.g., SA-VQE
response terms.

\subsection{Fermion-to-Qubit Operator Mapping}

Within any hybrid quantum/classical approach for the electronic Schr\"odinger
equation, we require a mapping between the spin orbitals and fermionic
composition operators of the electronic structure problem to the qubits and
distinguishable two-level composition operators of the quantum computer. There
are myriad approaches for this mapping, including (variants of) the Jordan-Wigner, parity, and
Bravyi-Kitaev maps. For concreteness, we use the Jordan-Wigner mapping in the
present work, as previously outlined by many authors and as previously used
within our local quantum-number-preserving (QNP) ansatz for MC-VQE in
\cite{QNP2021}. This
leads directly to the rather verbose representation of the Hamiltonian and
quantum number operators as specific linear combinations of Pauli operators in
the qubit basis, as detailed in Appendix~\ref{sec:JW}. Many techniques have
been developed to reduce the verbosity of the representation, notably including
the double factorization approach \cite{
poulin2014trotter,
peng2017highly,
motta2018low,
motta2019efficient,
kivlichan2018quantum,
berry2019qubitization,
matsuzawa2020jastrow,
huggins2021efficient,
cohn2021quantum},
composing the Hamiltonian as a sum of groups of simultaneously observable Pauli
operators, with each leaf in the sum corresponding to a specific spin-adapted
spatial orbital rotation obtained by tensor factorization of the ERIs. For our
purposes herein, it is conceptually sufficient to be able to compute the density
matrix in the natural representation of the qubit-basis operators, e.g., to be
able to compute the expectation value of each Pauli word for a representation of
the operators in terms of a linear combination of Pauli operators. In numerical
practice, it would surely be advantageous to use a more advanced representation
such as double factorization to reduce the measurement cost for each observable.
This could be seamlessly adopted to the framework presented here, with the
caveat that additional classical response terms might arise if an approximate
tensor factorization scheme is used to represent the Hamiltonian operator. We
defer such considerations to a later study.

\subsection{MC-VQE Active Space Wavefunctions}

We formally parametrize the MC-VQE active space wavefunctions as,
\begin{equation}
| \Psi^{\Theta} \rangle
\coloneqq
\sum_{\Theta'}
\hat U ( \{ \theta_{g} \} )
| \Phi^{\Theta'} \rangle
V_{\Theta' \Theta}
\coloneqq
\sum_{\Theta'}
| \Gamma^{\Theta'} \rangle
V_{\Theta' \Theta}
\end{equation}
Here $\{ | \Phi^{\Theta} \rangle \}$ are reference states which are classically
and quantumly efficiently describable and preparable. The specific construction
of $\{ | \Phi^{\Theta} \rangle \}$ is a user choice, subject to the following
requirements: (1) the reference states must have definite and proper
$N_{\alpha}$, $N_{\beta}$, and $S$ quantum numbers and (2) the procedure used to
define the reference states must either contain no auxiliary parameters or must
be describable in terms of the solutions of a set of nonlinear equations.
For concreteness, herein we choose these reference states to be configuration
state functions (CSFs), and they therefore have no additional internal
computational parameters. If reference states with auxiliary internal parameters
are used, additional quantum response equations may arise below, as occurred when
we used configuration interaction singles (CIS) reference states instead of CSFs
in our AIEM work.

$\hat U(\{ \theta_g \})$ is a VQE entangler circuit with quantum circuit
parameters $\{ \theta_g \}$. The specific construction of $\hat U(\{ \theta_{g}
\})$ is a user choice, subject to the following two requirements: (1) The
circuit operator $\hat U(\{ \theta_g \})$ must commute with the $\hat
N_{\alpha}$, $\hat N_{\beta}$, and $\hat S^2$ quantum number operators for all
parameter choices $\{ \theta_{g} \}$ and (2) The parameters of the circuit must
be simultaneously optimized as described below.  For concreteness, herein we use
the quantum number preserving gate fabric circuits defined in our prior work \cite{QNP2021},
though many other choices that satisfy (1) and (2) could be used within the
formalism derived below.

Once the construction of the VQE entangler circuit is established, the circuits
parameters $\{ \theta_{g} \}$ are optimized to minimize the state-averaged VQE
energy,
\begin{equation}
\theta_g^{*}
\coloneqq
\underset{
\theta_{g}
}{
\mathrm{argmin}
}
\underbrace{
\left (
\sum_{\Theta}
w_{\Theta}
\langle \Gamma^{\Theta} | 
\hat H
| \Gamma^{\Theta} \rangle
\right )
}_{\bar E (\{ \theta_{g} \})}
\end{equation}
Here $\{ w_{\Theta} \}$ are user-defined non-increasing state averaging
weights. In the above two equations, we define the ``entangled reference
states'' $| \Gamma^{\Theta} \rangle \coloneqq \hat U (\{ \theta_g \}) |
\Phi^{\Theta} \rangle$. $\hat H$ is the molecular Hamiltonian of the system, and
can be written in terms of second quantized operators or Pauli operators as
linear combinations of the one and two body integrals $(p|\hat h|q)$ and
$(pq|rs)$.

The weak form of the SA-VQE energy minimization condition is the first-order
stationary condition,
\begin{equation}
\pdiff{}{\theta_{g}}
\bar E
=
\pdiff{}{\theta_{g}}
\sum_{\Theta}
w_{\Theta}
\langle \Gamma^{\Theta} | 
\hat H
| \Gamma^{\Theta} \rangle
=
0
\end{equation}
In practice, this condition is used to define the SA-VQE entangler circuit
parameters $\{ \theta_{g} \}$.

The MC-VQE subspace eigenvectors $V_{\Theta' \Theta}$ are chosen to diagonalize
the MC-VQE subspace Hamiltonian
$\mathcal{H}_{\Theta \Theta'} \coloneqq \langle \Gamma^{\Theta} | \hat H |
\Gamma^{\Theta'} \rangle $,
\begin{equation}
\sum_{\Theta'}
\mathcal{H}_{\Theta \Theta'}
V_{\Theta' \Theta''}
=
V_{\Theta \Theta''}
E^{\Theta}
\end{equation}
additionally subject to the orthonormality constraints,
\begin{equation}
\sum_{\Theta''}
V_{\Theta \Theta''}
V_{\Theta' \Theta''}
=
\delta_{\Theta \Theta'}
,
\sum_{\Theta''}
V_{\Theta'' \Theta}
V_{\Theta'' \Theta'}
=
\delta_{\Theta \Theta'} 
\end{equation}
and the MC-VQE state energies $E^{\Theta}$ satisfy,
\begin{equation}
E^{\Theta}
=
\langle \Psi^{\Theta} | \hat H | \Psi^{\Theta} \rangle .
\end{equation}
Note that it is sometimes convenient to define the rotated reference state,
\begin{equation}
| \Omega^{\Theta} \rangle
\equiv
\sum_{\Theta'}
V_{\Theta' \Theta}
| \Phi^{\Theta} \rangle
:
\
| \Psi^{\Theta} \rangle
=
\hat U (\{ \theta_g \})
| \Omega^{\Theta} \rangle
\end{equation}
As a linear combination of classical reference states, this is usually an
efficient classically and quantumly representable state. The utility of this
rotated reference state is that a given MC-VQE eigenstate can be prepared with a
single quantum circuit, and therefore observables, derivatives, and unrelaxed
properties can be evaluated from observations of a single quantum circuit, e.g.,
\begin{equation}
E^{\Theta}
=
\langle \Omega^{\Theta} |
\hat U^{\dagger} (\{ \theta_g \})
\hat H
\hat U (\{ \theta_g \})
| \Omega^{\Theta} \rangle
\end{equation}

% In practice, these MC-VQE states are usually defined to be active in only a strict
% active space of the FON-RHF orbitals, i.e., outside this subspace they are taken to be
% either completely filled or empty. This provides for reduced quantum circuit
% resource requirements, and lowers the cost of forming certain classical
% computational intermediates, but does not affect the structure of the
% FON-RHF-MC-VQE Lagrangian (i.e., all of the same terms would be present even if
% all FON-RHF orbitals were included in the active space).

All in all, MC-VQE can be thought of as a procedure that adjusts the parameters
of the entangler circuit such that the subspace spanned by the chosen reference
states is rotated into the lowest energy corner of the Hilbert space reachable with the
chosen entangler ansatz. The Hamiltonian in this subspace can then be measured
and classically diagonalized. This yields variational ground and excited state energies and also recipes to prepare explicitly orthogonal corresponding wave functions, without
having to enforce this orthogonality via measurements of overlaps or higher 
powers of the Hamiltonian. Even if one is not interested in excited state properties
MC-VQE can be be an attractive method because its multireference nature also means
that its ground state energy may be lower than the energy of any 
individual state that is preparable via the entangler ansatz from a single reference state.

\subsection{MC-VQE Active Space Lagrangian}

The ``quantum'' part of the Lagrangian is,
\begin{equation}
\mathcal{L}^{\Theta}
\equiv
E^{\Theta}
+
\sum_{\theta_{g}}
\pdiff{\bar E}{\theta_{g}}
\lambda_{g}
\end{equation}
Here $\{ \lambda_{g} \}$ are Lagrange multipliers that enforce the use of the
SA-VQE stationary conditions to choose the VQE entangler circuit parameters $\{
\theta_{g} \}$,
\begin{equation}
\pdiff{\mathcal{L}^{\Theta}}{\lambda_{g}}
=
0
\Rightarrow
\pdiff{\bar E}{\theta_{g}}
=
0
\end{equation}

Note that one may additionally choose to add terms corresponding to the subspace
eigendecomposition (i.e., Lagrangian terms to define $V_{\Theta \Theta'}$ as
diagonalizing the subspace Hamiltonian and being orthonormal), but the response
terms stemming from these contributions are zero due to the variational nature
of the eigendecomposition.

\subsection{MC-VQE Active Space Gradient}

The target of the ``quantum'' part of the gradient are the spin-summed relaxed
OPDM and TPDM [first defined in Eqs.~\eqref{eq:relaxed_opdm} and
\eqref{eq:relaxed_tpdm}]
in the active spatial orbitals, for which it holds that
\begin{align}
\bar \gamma_{pq}^{\Theta}
&=
\diff{E^{\Theta}}{(p|\hat h|q)}
=
\pdiff{\mathcal{L}^{\Theta}}{(p|\hat h|q)}
\intertext{and}
\bar \Gamma_{pqrs}^{\Theta}
&=
2
\diff{E^{\Theta}}{(pq|rs)}
=
2
\pdiff{\mathcal{L}^{\Theta}}{(pq|rs)} .
\end{align}
Due to the properties of the Lagrangian formalism, we do not need to compute the
parameter response portions of the following chain rule derivatives explicitly,
\begin{align}
\bar \gamma_{pq}^{\Theta}
&=
\sum_{g}
\pdiff{E^{\Theta}}{\theta_{g}}
\pdiff{\theta_{g}}{(p | \hat h | q)}
\intertext{and}
\bar \Gamma_{pqrs}^{\Theta}
&=
2
\sum_{g}
\pdiff{E^{\Theta}}{\theta_{g}}
\pdiff{\theta_{g}}{(pq|rs)}
\end{align}
Instead, the properties of the Lagrangian formalism allow us to compute, e.g.,
\begin{align}
\bar \gamma_{pq}^{\Theta}
&=
\pdiff{E^{\Theta}}{(p|\hat h|q)}
+
\sum_{g}
\pdiff{^2 \bar E}{(p|\hat h|q) \partial \theta_{g}}
\lambda_{g}
\\
&\equiv
\gamma_{pq}^{\Theta}
+
\tilde \gamma_{pq}^{\Theta}
\end{align}
directly. The last line shows the popular partition of the density matrix into unrelaxed
and response contributions.

\subsection{Quantum Observables Required for MC-VQE Energies and Gradients}

Inspection of the above reveals that the following classes of quantum
observables are needed to compute MC-VQE energies and gradients:

(1) Diagonal Expectation Values:
\begin{equation}
\mathcal{O}^{\Theta}
=
\langle \Phi^{\Theta} |
\hat U^{\dagger} (\{ \theta_g \})
\hat O
\hat U^{\dagger} (\{ \theta_g \})
| \Phi^{\Theta} \rangle
\end{equation}
Here $\hat O$ is usually the Hamiltonian, but could optionally be a quantum
number operator or any other symmetric operator. $| \Phi^{\Theta} \rangle$ can
be either a reference state or a rotated reference state.

(2) Off-Diagonal Expectation Values:
\begin{equation}
\mathcal{O}^{\Theta \Theta'}
=
\langle \Phi^{\Theta} |
\hat U^{\dagger} (\{ \theta_g \})
\hat O
\hat U^{\dagger} (\{ \theta_g \})
| \Phi^{\Theta'} \rangle
\end{equation}
\[
=
\langle \chi_{+}^{\Theta \Theta'} |
\hat U^{\dagger} (\{ \theta_g \})
\hat O
\hat U^{\dagger} (\{ \theta_g \})
| \chi_{+}^{\Theta \Theta'} \rangle
/ 2
\]
\[
-
\langle \chi_{-}^{\Theta \Theta'} |
\hat U^{\dagger} (\{ \theta_g \})
\hat O
\hat U^{\dagger} (\{ \theta_g \})
| \chi_{-}^{\Theta \Theta'} \rangle
 / 2
\]
Where the classically and quantumly tractable interfering combinations of
reference states are,
\begin{equation}
| \chi_{\pm}^{\Theta \Theta'} \rangle
\equiv
\frac{1}{\sqrt{2}}
\left [
| \Phi^{\Theta} \rangle
\pm
| \Phi^{\Theta'} \rangle
\right ]
\end{equation}

(3) Diagonal Density Matrix Expectation Values:
\begin{equation}
\gamma_{I}
\equiv
\pdiff{\mathcal{O}^{\Theta}}{O_{I}}
=
\langle \Phi^{\Theta} |
\hat U^{\dagger} (\{ \theta_g \})
\hat \Pi_{I}
\hat U^{\dagger} (\{ \theta_g \})
| \Phi^{\Theta} \rangle
\end{equation}
For cases where the operator $\hat O$ is composed as a weighted linear
combination of ``basis operators.''
\begin{equation}
\hat O 
\equiv
\sum_{I}
O_{I}
\hat \Pi_{I}
\end{equation}
Here the basis operators $\{ \hat \Pi_{I} \}$ could be, e.g., Pauli words,
spin-summed single substitution operators $\hat E_{pq}^{+}$, etc. Note in
particular that one can classically obtain the density matrix in terms of
spin-summed single substitution operators (or products thereof) from the
corresponding density matrix in terms of Pauli words if the details of the
mapping of the second quantized composition operators to Pauli words is known.
We refer to this process as ``backtransforming'' the density matrix, and refer
the reader to Appendix~\ref{sec:JW} for more details on the procedure. This
procedure can also be adapted to the case where the density matrix is
expressed in double factorized representation instead of Pauli word
representation.  

(4) Diagonal Parameter Gradients:
\begin{equation}
\pdiff{\mathcal{O}^{\Theta}}{\theta_{g}}
=
\sum_{P}
v_{P}
\mathcal{O}^{\Theta} (\{ \theta_{g'} + \delta_{gg'} t_{P} \})
\end{equation}
This usually refers to the case where $\{ \theta_{g} \}$ are the parameters of
the SA-VQE entangler circuit. The equality shows the analytical expression of
parameter gradients for quantum circuits in terms of the well-known parameter
shift rule or one of its generalizations \cite{PhysRevLett.118.150503,PhysRevA.98.032309,PennyLane, Mari2021,wierichs2021general}.
Within such rules the derivative of the expectation value with
respect to each circuit parameter can be analytically computed as a weighted
linear combination of separate observable expectation values evaluated along a
widely spaced stencil of displacements of that parameter. The fact that the
displacements are widely spaced (e.g., not finite difference) implies that
statistical noise does not specifically magnify in this process. The parameter
shift displacements and weights $\{ < t_{P} , v_{P} > \}$ are specific to the
eigenstructure of each gate element, and in particular to the corresponding
number of unique frequencies in the trigonometric polynomial of the observable
tomography along the gate parameter. E.g., simple $\hat R_{y} (\theta)$ gates
require a 2-point parameter shift rule, a diagonal pair substitution gate
$\mathrm{QNP_{PX}} (\theta)$ requires a 4-point parameter shift rule, and a
spin-adapted orbital rotation $\mathrm{QNP_{OR}} (\theta)$ requires an 8-point
parameter shift rule \cite{QNP2021}. Two notes of emphasis are appropriate here for contrast
with the later section on approximate parameter shift rules for directional
derivatives: (1) for gradients evaluated along single parameter directions, the
parameter shift rule is exact with a \emph{constant} number of stencil points
but (2) as soon as a directional derivative that involves a linear combinations
of parameter directions is involved, it appears that an exact parameter shift
rule requires a \emph{linear} number of stencil points (essentially computing
the full gradient with the original parameter shift rule, and then classically
contracting with the desired direction vector).

(5) Diagonal Parameter Hessians:
\begin{equation}
\pdiff{^2\mathcal{O}^{\Theta}}{\theta_{g} \partial \theta_{g'}}
=
\sum_{P}
\sum_{Q}
v_{P}
v_{Q}
\mathcal{O}^{\Theta} (\{ \theta_{g''} + \delta_{gg''} t_{P} + \delta_{g'g''}
t_{Q}\})
\end{equation}
I.e., a double application of the parameter shift rule. It  should be noted that evaluation of the diagonal elements of the Hessian can be obtained in essentially the same linear-scaling effort as the gradient, but the evaluation of the full Hessian requires quadratic-scaling effort \cite{Mari2021,wierichs2021general}. Below we consider an approach to try to reduce the scaling of computations involving the Hessian by considering a Hessian matrix-vector product formalism in concert with widely-spaced finite difference stencils for directional derivatives. The connection between finite difference and parameter-shift gradients have recently been discussed in \cite{Mari2021,Hubregtsen2021}, albeit in the context of single parameter gradients and not directional derivatives.
Note that off-diagonal density matrix elements and/or observable derivatives are
also easily obtained by a combination of appropriate rule and the off-diagonal
observable expectation value rule.

\subsection{SA-VQE Response Equations}

In order to obtain the target relaxed density matrices of the quantum gradient,
we must first solve the SA-VQE response equations,
\begin{equation}
\pdiff{\mathcal{L}^{\Theta}}{\theta_{g'}}
=
0
\Rightarrow
\sum_{g}
\pdiff{^2 \bar E}{\theta_{g'} \partial \theta_{g}}
\lambda_{g}
=
-
\pdiff{E^{\Theta}}{\theta_{g}}
\end{equation}
As written, the SA-VQE response equations appear to require evaluation and
storage of the $\mathcal{O} (N_{\theta}^2)$ SA-VQE Hessian matrix,
\begin{equation}
\mathcal{A}_{gg'}
\equiv
\pdiff{^2 \bar E}{\theta_{g'} \partial \theta_{g}}
\end{equation}

Note that there are a few important limits in which the RHS of the SA-VQE
response equations,
\begin{equation}
b_{g} \equiv 
-
\pdiff{E^{\Theta}}{\theta_{g}}
\end{equation}
will be zero and where we can avoid the solution of the SA-VQE response
equations: (1) In the case that we only include a single state in the SA-VQE
optimization, i.e., using VQE instead of MC-VQE. In this case, the state
averaged gradient that is optimized is equivalent to the state-specific RHS
gradient, and is therefore zero at the conclusion of the VQE parameter
optimization. (2) In the limit that we use a powerful enough entangler circuit
to exactly solve the Schr\"odinger equation for the targeted states. (3) In the
limit that we are not at the complete entangler circuit limit, but where the
state-specific RHS gradient is accidentally zero for all states sought due to
external considerations such as the targeted states being in different spatial
symmetry irreps.

\subsection{Iterative Solution of the SA-VQE Response Equations}

\label{sec:sa_vqe_response_diis}

Over the next two sections we will introduce an approach to remove the explicit
need for the generation, storage, and manipulation of the SA-VQE Hessian. The
aim is to remove the need for a double application of the parameter shift rule,
which currently scales as $\mathcal{O}(N_{\theta}^2)$. This is formally
higher scaling than the single application of the parameter shift rule needed
for the SA-VQE energy gradient, $\mathcal{O}(N_{\theta})$, i.e., as needed in
gradient-based optimization of the SA-VQE entangler circuit parameters. Put
another way, the computation of the gradient currently seems to scale higher
[$\mathcal{O}(N_{\theta}^2)$] than the computation of the energy
[$\mathcal{O}(N_{\theta})$] due to the cost of evaluating the SA-VQE Hessian
before the solution of the response equations. In most classical electronic
structure methods, the same scaling can be achieved in both the energy and the
gradient. The following procedure is one approach to realizing this scaling
reduction for SA-VQE response:

The first step needed in this scaling reduction is to replace the explicit
solution of the SA-VQE response equations with an iterative method based on
matrix-vector products, e.g., a Krylov-type method. There are many possible
and common choices to accomplish this, including but not limited to GMRES,
MINRES, BiCGSTAB, SOR, Jacobi, or Gauss-Seidel approaches. Note that we cannot
use standard conjugate gradient (CG or PCG) methods here, as the SA-VQE Hessian
might be indefinite for numerical reasons.  Here we focus on the particularly
simple choice of a fixed-point iteration accelerated by Pulay's Direct Inversion
of the Iterative Subspace (DIIS) extrapolation method \cite{Pulay:1980:393,
Pulay:1982:556}, as is often used in the solution of response equations (and
also nonlinear parameter optimization equations) in classical electronic
structure methods.

The specific procedure is as follows for solving $\hat A \vec \lambda = \vec b$,
with a symmetric indefinite linear operator
$\hat A$ (the SA-VQE Hessian), a RHS
of $\vec b$ (the negative to the state-specific VQE parameter gradient), a LHS
of $\vec \lambda$ (the target Lagrangian vector), and an easily obtained and
pseudoinverted symmetric preconditioner $\hat P$ that approximates $\hat A$ in a
spectral sense:
\begin{enumerate}
\item Initialize the solution vector to the trivial initial guess $\vec \lambda
= \vec 0$.
\item Compute the residual $\vec r \equiv \vec b - \hat A \vec \lambda$.
\item If the maximum residual is less than a user specified convergence
threshold $\delta$, i.e., if $\| \vec r \|_{\infty} < \delta$, exit the
procedure with the current $\vec \lambda$ as the converged Lagrangian solution
vector.
\item Compute the preconditions residual $\vec d \equiv \hat P^{-1} \vec r$. 
\item Compute the updated solution vector $\vec \lambda \leftarrow \vec \lambda
+ \vec d$.
\item Add the current values of the state vector $\vec \lambda$ and error vector
$\vec r$ to the DIIS history (note that some implementations may prefer to use
$\vec d$ as the error vector).
\item Query the DIIS protocol for an extrapolated guess to the solution vector
$\vec \lambda$.
\item Repeat from Step 2 until convergence is achieved in Step 3 or a maximum
number of iterations is reached and the iterative procedure reports failure.
\end{enumerate}
Note the key dependence on the SA-VQE Hessian matrix-vector product operation
$\vec \sigma ( \vec x) \equiv \hat A \vec x$ in Step 2.
For the preconditioner, herein we use the diagonal of the Hessian (obtainable in
$\mathcal{O}(N_{\Theta})$ scaling via the double parameter shift rule
\cite{PhysRevLett.118.150503,PhysRevA.98.032309,PennyLane,
Mari2021,wierichs2021general} as sketched for simpler entangler circuits in our
previous derivative work \cite{parrish2019hybrid}),
conditioned to remove elements with extremely small magnitude. The use of
more-advanced preconditioners to improve convergence is an important topic for
future research.

The well-known DIIS extrapolation protocol holds a limited history of state and
error vector pairs. At each step in the fixed point iteration, the DIIS
procedure builds a linear error model, and solves norm-constrained least-squares
equations to provide an extrapolated prediction of the converged limit of the
fixed-point iteration. The DIIS protocol is easily constructed with a handful of
classical linear algebra operations, and can be made to be completely agnostic
of the details of the fixed point iteration being considered. DIIS can solve
nonlinear systems equations effectively in many cases, and when applied to
linear indefinite systems of equations in the manner described above, has been
shown to be a form of GMRES.  For more information on DIIS, the reader is
referred to a number of articles on the classical use of DIIS in classical
electronic structure methods
\cite{Pulay:1980:393,Pulay:1982:556,csaszar1984geometry,hamilton1986direct,hutter1994electronic,scuseria1986accelerating,kudin2002black,chen2011listb,hu2017projected}, as well as our hybrid quantum/classical Jacobi
parameter optimization approach which uses DIIS as a convergence accelerator
\cite{parrish2019jacobi}. A complete numerical recipe for the DIIS protocol is
laid out in the last of these papers.

\subsection{Finite-Difference Approximation of SA-VQE Hessian-Vector Products}

\label{sec:sa_vqe_response_fd}

Now that we have transformed the SA-VQE response equations into a matrix-vector
product formalism, we have a chance at reducing the scaling by considering the
specifics of the SA-VQE Hessian-vector product, shown here for an arbitrary
trial vector $x_{g}$,
\begin{equation}
\sigma_{g} [x_{g'}]
\equiv
\sum_{g'}
\mathcal{A}_{gg'}
x_{g'}
=
\sum_{g'}
\pdiff{^2\bar E}{\theta_{g} \partial \theta_{g'}}
x_{g'}
\end{equation}
Considering the specific normalized linear combination of parameters along an
auxiliary parameter $\tilde t$,
\begin{equation}
\theta_{g'} (\tilde t)
\equiv
\theta_{g'}^{0}
+
\frac{x_{g'}}{\| \vec x \|_2}
\tilde t
\end{equation}
Then our task is to compute the gradient of the scaled directional derivative of
the state-averaged energy in the direction of the trial vector,
\begin{equation}
\sigma_{g} [x_{g'}]
=
\pdiff{}{\theta_{g}}
\pdiff{\bar E}{\tilde t}
\| \vec x \|_2
\end{equation}
It would be ideal to express the directional derivative $\partial_{\tilde t} \bar E$ as
a widely-spaced parameter-shift-rule-type stencil in $t$,
\begin{equation}
\pdiff{\bar E}{\tilde t}
\stackrel{?}{\approx}
\sum_{P}
v_{P}
\bar E (\{ \theta_{g'}^{0} + \tilde t_{P} x_{g}' / \| \vec x \|_2\})
\end{equation}
Here $\{ < \tilde t_{P}, \tilde v_{P} > \}$ are the points and weights of the
parameter-shift-rule-type stencil. If such a stencil is possible with a constant
number of points for the directional derivative, the complete SA-VQE Hessian
matrix-vector product could be formed in $\mathcal{O}(N_{\theta})$ effort by
then applying the usual first-derivative parameter shift rule $\{ < t_Q, v_Q >
\}$ to each point in the directional derivative stencil to obtain the second
derivative in $\theta_{g}$,
\begin{equation}
\sigma_{g} [x_{g'}]
\approx
\sum_{Q}
\sum_{P}
v_{Q}
\tilde v_{P}
\bar E(\{ \theta_{g'}^{0} + \tilde t_{P} x_{g'} / \| \vec x \|_2 + \delta_{gg'}
t_{Q} \})
\| \vec x \|_2
\end{equation}
I.e., for each needed SA-VQE Hessian matrix-vector product, we first parameter
shift in a constant-scaling stencil along the normal direction of the trial
vector $\vec x$, and then perform a second parameter shift along each parameter
direction to obtain the needed second derivative vector. The second shift can be
performed with the usual exact parameter shift rule appropriate for the
parameter/gate in question. The first proposed parameter shift rule along the
arbitrary trial vector direction does not appear to have an exact solution in
less than exponential effort, due to the exponential number of combinations of
trigonometric functions in an arbitrary direction in the
$N_{\theta}$-dimensional trigonometric landscape of $E ( \{ \theta_{g} \})$.
However, we argue that it is plausible that a widely spaced-finite-difference
stencil in $\tilde t$ may provide for a highly accurate approximate computation
of the directional derivative due to the directionally bandlimited nature of the 
$N_{\theta}$-dimensional trigonometric landscape of $E ( \{ \theta_{g} \})$. In
particular, if we consider the case of gates all with maximum angular frequency
oscillation (i.e., trigonometric polynomial basis function) in parameter space
in parameter space of $M (\theta) = \exp(-i \omega \theta)$, where $\omega$ is
the maximum angular frequency, then the cut through the parameter space along
$\tilde t$ is composed of a linear combination of an exponential number of
trigonometric polynomial basis functions, but with a maximum angular frequency
oscillation of $M' (t) = \exp(-i \omega \| \vec x \|_1 / \| \vec x \|_2)$. I.e.,
the maximum angular frequency along $\tilde t$ is $\omega \| \vec x \|_1 / \|
\vec x \|_2 \leq \omega \sqrt{N_{\theta}}$. The worst case is achieved for the
major diagonal trial vector $x_{g} = 1 \ \forall \ g$. 
 Overall, the $\sqrt{N_{\theta}}$ worst-case scaling of the maximum frequency of
$\bar E (\tilde t)$ indicates that the function is significantly bandlimited
along any direction. This leads to the ability to apply widely-spaced finite
difference stencils without significant loss of accuracy. 

In future work, it may be worth considering an adaptive form of finite
difference stencil that changes the step size as a function of $\| \vec x \|_1 /
\| \vec x \|_2$. However, we have empirically found that the application of a
simple Newton-Cotes symmetric finite difference stencil with a large and
isotropic step size for all $\vec x$ works surprisingly well for this
application in the numerical examples encountered below.

\subsection{Procedure for MC-VQE Energies and Gradients}

For clarity, we provide this section to explicitly enumerate the full
time-ordered procedure for MC-VQE energies and gradients. Many of these
operations have been discussed mathematically above, but the time order and
preferred approach for each required quantum observable are highlighted in this
section. 

MC-VQE Energies:
\begin{enumerate}
\item Classically obtain a description of the active space spatial orbitals $\{
\phi_{p} (\vec r_1) \}$.
\item Classically obtain the active space matrix elements $E_{\mathrm{ext}}$,
$\{ (p|\hat h|q) \}$, and $\{ (pq|rs) \}$.
\item Determine the details of the fermion-to-qubit mapping of the composition
operators, Hamiltonian operator, and quantum number operators.
\item Construct the quantum number pure reference states $\{ | \Phi^{\Theta} \rangle \}$ for the
$N_{\Theta}$ targeted states and for the target quantum numbers $N_{\alpha}$,
$N_{\beta}$, and $S$ (the quantum numbers may vary from state to state, though this is not
explicitly considered in this work). This
included classically determining the ansatz any internal
parameters of the reference states, and also designing quantum circuits to
efficiently prepare these reference states from the qubit fiducial state $|\vec
0\rangle$.
\item Construct the conceptual design of a quantum number preserving SA-VQE
entangler circuit $\hat U ( \{ \theta_{g} \})$, including gate layout and
composition and initial parameter guess.
\item Optimize the SA-VQE energy $\bar E(\{ \theta_{g} \})$ with user-supplied
state averaging weights $\{ w_{\Theta} \}$ with respect to the
entangler circuit parameters $\{ \theta_{g} \}$. There are myriad iterative
approaches for this parameter optimization which may rely on quantum expectation
values of the Hamiltonian over the diagonal entangled reference states $\bar
E(\{ \theta_{g} \}) \equiv \sum_{\Theta} w_{\Theta} \langle \Gamma^{\Theta} |
\hat H | \Gamma^{\Theta} \rangle$, and/or gradients thereof
$\partial_{\theta_{g}} \bar E$  as computed by the parameter shift rule.
\item Compute the off-diagonal elements of the MC-VQE subspace Hamiltonian
$\{ \mathcal{H}_{\Theta \Theta'} \equiv \langle \Gamma^{\Theta} | \hat H |
\Gamma^{\Theta'} \rangle \}$ by using differences of Hamiltonian expectation values
over entangled interfering combinations of reference states.
\item Classically diagonalize $\mathcal{H}_{\Theta \Theta'}$ to obtain the
eigenvectors $V_{\Theta' \Theta}$ and the corresponding eigenvalues
$E^{\Theta}$. The diagonal elements of the subspace Hamiltonian will likely
already be available from the SA-VQE energy computation in the previous
parameter optimization step. Note that the entangled reference states are
defined as $\{ |\Gamma^{\Theta} (\{ \theta_{g} \}) \rangle \equiv \hat U
(\{\theta_{g} \}) | \Phi^{\Theta} \rangle \}$.
\item Optionally classically compute the rotated reference states $\{ |
\Omega^{\Theta} \rangle \equiv \sum_{\Theta'} V_{\Theta' \Theta} |
\Phi^{\Theta'} \rangle \}$, including classical composition and description of
corresponding quantum circuit.
\end{enumerate}

MC-VQE Gradients:
\begin{enumerate}
\item Compute the state-specific VQE entangler parameter gradient $
\partial_{\theta_{g}} E^{\Theta}$, e.g., using the parameter shift rule in
concert with Hamiltonian expectation values over the entangled rotated reference
state for state $\Theta$.
\item Solve the SA-VQE response equations for RHS of $b_{g} \equiv -
\partial_{\theta_{g}} E^{\Theta}$ to obtain the Lagrangian $\lambda_{g}$. This
can be accomplished by any combination of the explicit, iterative, or iterative
+ finite difference SA-VQE Hessian-vector product methods described in Sections
\ref{sec:sa_vqe_response_diis} or \ref{sec:sa_vqe_response_fd} above.
\item Compute the unrelaxed state-specific OPDM and TPDM $\gamma_{pq}^{\Theta}
\equiv \partial_{(p|\hat h|q)} E^{\Theta}$ and $\Gamma_{pqrs}^{\Theta} \equiv
\partial_{(pq|rs)} E^{\Theta}$, e.g., using Pauli word expectation values over
the entangled rotated reference state for state $\Theta$, followed by
backtransformation of the Pauli word expectation values to the spin-summed
spatial orbital density matrices.
\item Compute the SA-VQE response contributions to the OPDM and TPDM, e.g.,
$\tilde \gamma_{pq}^{\Theta} \equiv \sum_{g} [\partial_{\theta_{g}}
\partial_{(p|\hat h|q)} \bar E] \lambda_{g}$. This can be done efficiently with
a combination of the Pauli-word-type procedure for the unrelaxed OPDM/TPDM and a
single application of the parameter shift rule.
\item Add the unrelaxed and SA-VQE response contributions to form the
``quantum'' relaxed OPDM and TPDM, e.g., $\bar \gamma_{pq}^{\Theta} \equiv
\gamma_{pq}^{\Theta} + \tilde \gamma_{pq}^{\Theta}$.
\item Send the relaxed quantum OPDM and TPDM to the classical electronic
structure code to determine the classical response (e.g., orbital response),
basis set (e.g., Pulay term), and intrinsic contributions to the gradient.
Formally this computes $\mathrm{d}_{\xi} E^{\theta} = \mathrm{d}_{\xi}
E_{\mathrm{ext}} + \sum_{pq} \bar \gamma_{pq}^{\Theta} \mathrm{d}_{\xi} (p| \hat
h|q) + \frac{1}{2} \sum_{pqrs} \bar \Gamma_{pqrs}^{\Theta} \mathrm{d}_{\xi}
(pq|rs)$.
\end{enumerate}

Note that double factorization or other Hamiltonian compression approaches could
be substituted anywhere Hamiltonian expectation values or Pauli-to-orbital
density matrix computations are encountered.

\section{Computational Details}

For the classical orbital determination, Hamiltonian matrix element evaluation,
and gradient evaluation steps, we used RHF and FON-RHF methods in
atom-centered Gaussian atomic orbital basis sets as implemented in the
\textsc{Lightspeed} + \textsc{TeraChem} code stack.  The quantum portions of
the MC-VQE energy and gradient approaches defined above were implemented in two 
independent code stacks; An in-house C++/Python quantum circuit simulator code. This code is capable of
evaluating Hamiltonian expectation values via either Pauli expectation value
expansion or by direct application of the Hamiltonian to an arbitrary qubit
statevector with known $N_{\alpha}$ and $N_{\beta}$ via a Knowles-Handy
matrix-vector product method.  This is similar in spirit to other
implementations proposed recently \cite{rubin2021fermionic,stair2021qforte}, though we choose to retain
the full $2^{N}$ statevector memory during the computation of the quantum
circuit simulation to allow for arbitrary gate decompositions (including
intermediate non-number-preserving gates) to be simulated. 
A code stack leveraging PennyLane \cite{PennyLane} for automatic differentiation and
and OpenFermion \cite{OpenFermion} for Hamiltonian transformations.
Both implementations were cross-checked against each other.
As the focus of the
present work is to enumerate the complete response-including analytical
gradient and to probe the size of response terms for realistic systems, we
perform the quantum circuit simulations in the limit of infinite statistical
sampling and without decoherence noise channels. Consideration of shot and
decoherence noise within the complete MC-VQE energy + gradient workflow is an
important topic that we leave for future study.

\section{Results}

\subsection{Moderate-Scale Test Case}

To validate our analytical gradient methodology within a challenging and
physically relevant system, we consider the case of cyclohexadiene
(specifically cyclohexa-1-3-diene) near one of its minimal energy conical
intersections (MECI). Cyclohexadiene exhibits an interesting electrocyclic ring
opening reaction with strong enantomeric selectivity upon photoexcitation with
UV light near 267 nm. This photochemical reaction is likely representative of
general electrocyclic ring-opening reactions in generic 1-3-cyclodiene
compounds such as terpenine, pro-vitamin-D3, and many others.  Understanding of
the non-adiabatic dynamics of this reaction, including yields, timescales,
stereochemistry  and mechanisms involves significant efforts on both the
dynamics and electronic structure.

Herein, we take the geometry of the ``ring-closing'' MECI from
SA-2-$\alpha$-CASSCF(6e, 4o)/6-31G* from Ref \cite{wolf2019photochemical}. We
then treat the system with FOMO-CASCI(6e, 4o, $\beta=0.3$)/6-31G* (Gaussian
blurring and same active space and $N_{\mathrm{FOMO}}$ as forthcoming CASCI
during the FON-RHF procedure). As FON-RHF and SA-$\alpha$-CASSCF use
different metrics for determining the orbitals, the FOMO-CASCI $S_0$ and $S_1$
states are not exactly at a conical intersection - this is useful as it
provides a sort of ``canonicalization'' of the $S_0$ and $S_1$ states and
allows us to robustly consider FCI vs. MC-VQE treatments within the active
space without considering degenerate rotations between states. 

\begin{figure}[ht!]
\begin{center}
\includegraphics[width=2.5in]{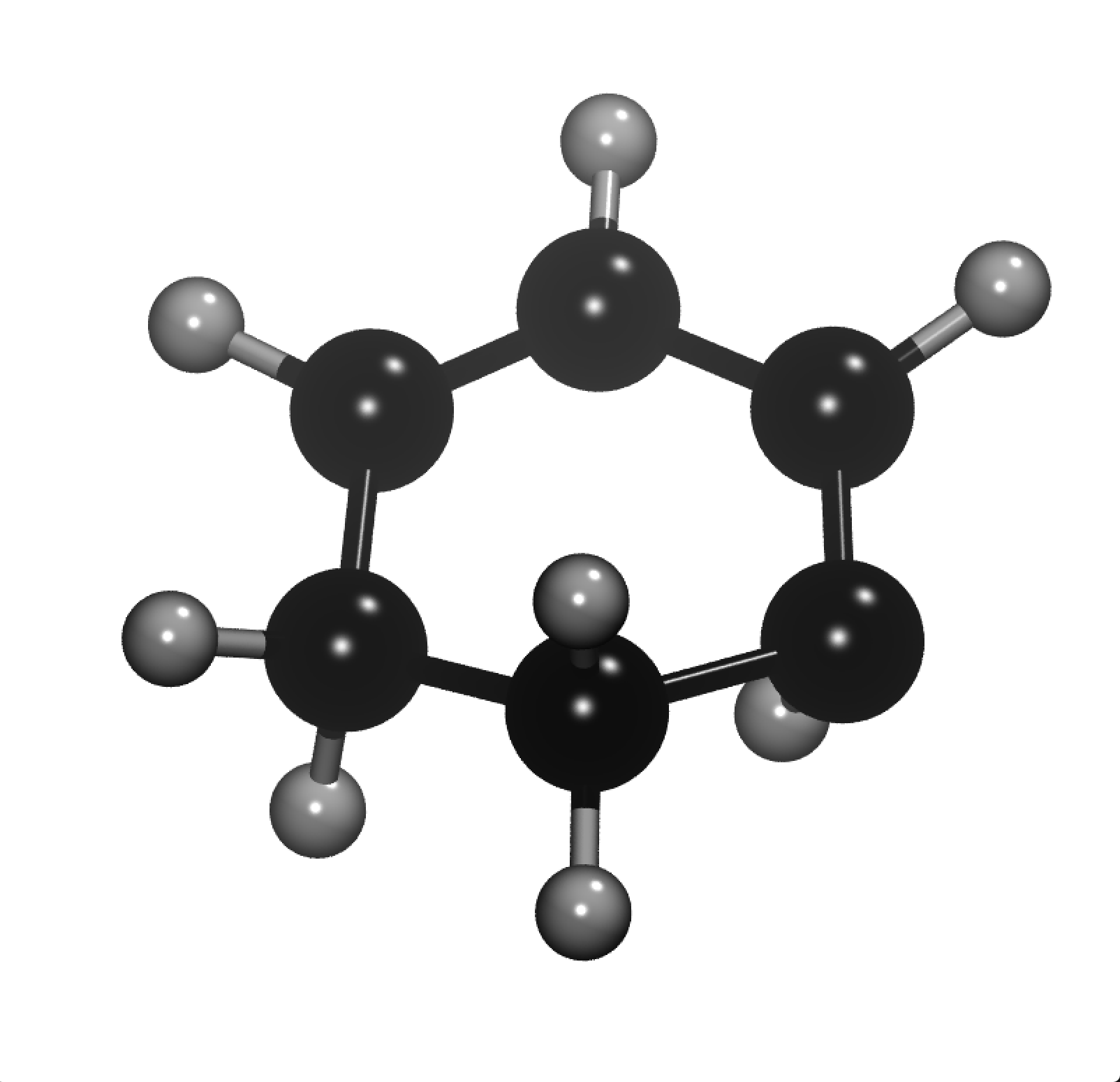}
\caption{Geometry of ``ring-closing'' cyclohexadiene MECI computed at
SA-2-$\alpha$-CASSCF(6e, 4o)/6-31G* in Ref. \cite{wolf2019photochemical} and
used as a test case within the present manuscript. The lower-right C-H bond is
distorted well out of plane, bringing $S_0$ and $S_1$ states into coincidence
without promoting ring-opening photochemistry. A different MECI is responsible
for the ring-opening channel of the dynamics.}
\label{fig:chd-geom}
\end{center}
\end{figure}

\subsection{Finite Difference Validation}

To validate the correctness of the derived and implemented methodology, we
performed finite difference computations with and without SA-VQE response, with
finite difference perturbations performed at the level of second-quantized
Hamiltonian matrix elements $(p|\hat h|q)$ and $(pq|rs)$ and at the level of
nuclear gradient computations. These computations indicate that the complete
gradient, including SA-VQE response, is concordant with the analytical gradient. 

For instance, for the MC-VQE nuclear gradient of the ground state for the
cyclohexadiene system considered above, with one ``double layer'' of QNP
entangler gates (one complete layer of even- and odd-pairing QNP gates between
spatial orbitals) and two states considered in SA-VQE, the maximum absolute
deviation (defined for two nuclear gradients $\vec G_1$ and $\vec G_2$  as
$M\equiv \|
\vec G_1 - \vec G_1 \|_{\infty}$) between the analytical or finite difference
CASCI and MC-VQE gradients is $0.016$ a.u., reflecting the incompleteness of the
MC-VQE entanglers at this rather coarse level of theory. The maximum absolute deviation
between the bare (not including SA-VQE response) MC-VQE gradients and analytical
or finite difference MC-VQE gradients is $0.0016$ a.u., indicating that the bare
and response-including MC-VQE gradients agree internally to about $10\times$
better than the CASCI and MC-VQE gradients, but that there are still nontrivial
response contributions to the gradient. Finally, the SA-VQE-response-including
MC-VQE gradients agree with the finite difference MC-VQE gradients to a maximum
absolute deviation of $4.6\times10^{-5}$. This last $~\sim35\times$ reduction in
maximum absolute deviation reveals the need for the SA-VQE response
contributions to produce the analytical gradient. Further analysis indicates
that the conservative maximum magnitude of the response contribution for this
case is $\Gamma \equiv \| G_{\mathrm{VQE-Resp}} - G_{\mathrm{VQE-Bare}}
\|_{\infty} / \| G_{\mathrm{VQE-Resp}} \|_{\infty} = 0.02$, i.e., that the
response contribution for this case is $\sim 2\%$ of the maximum analytical
gradient entry. This contribution is small but already nontrivial, and will likely
become more important in larger cases where VQE entangler is further from
completeness, and/or in other derivative contributions that are less stationary
than the state energy, e.g., the polarizability or non-adiabatic coupling
vector. 

\subsection{SA-VQE Hessian Matrix-Vector Product Response Methodology}

To probe the potential utility of the iterative solution of the SA-VQE response
equations with finite-difference Hessian matrix-vector products and DIIS
acceleration, we consider the application of this method to the cyclohexadiene
test case with various finite difference stencil sizes $n_{\mathrm{FD}}$ ranging
from $2$ to $10$ by even integers and various large finite difference stencil
sizes $\Delta_{\mathrm{FD}} \in [0.05, 0.1, 0.2, 0.3]$. As a preliminary,
solving the SA-VQE response equations with DIIS without any approximation in the
matrix-vector product yields essentially double precision machine epsilon
agreement with the explicit pseudoinversion of the full SA-VQE response
equation, and requires 5 full iterations of DIIS to reach a convergence in the
SA-VQE residual of $\| \vec r \|_{\infty} < 10^{-9}$. 

\begin{figure}[ht!]
\begin{center}
\includegraphics[width=3.4in]{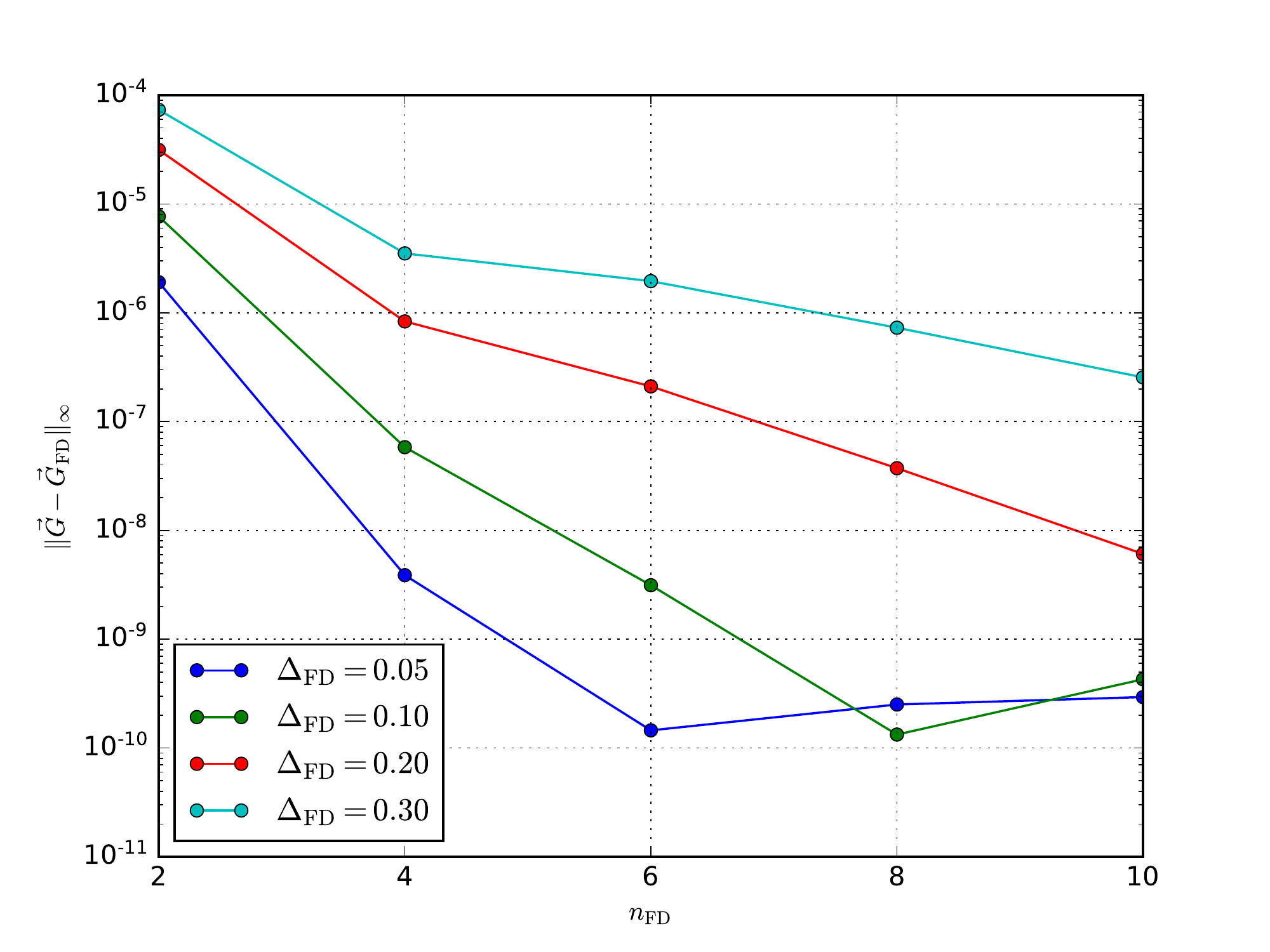}
\caption{Maximum absolute error in complete nuclear gradient for $S_0$ state of
cyclohexadiene ring-closing MECI computed with FOMO-CAS-MC-VQE(6e, 4o)/6-31G*
with 1 complete layer in the QNP entangler circuit. Here the error is shown as a
function of finite difference stencil sizes ($n_{\mathrm{FD}}$) and finite
difference step sizes ($\Delta_{\mathrm{FD}}$).}
\label{fig:diis_error}
\end{center}
\end{figure}

As Figure \ref{fig:diis_error} shows, the addition of finite difference
approximations in the matrix vector products does not appear to significantly
affect the accuracy of the resultant gradient, even with small $n_{\mathrm{FD}}$
or large $\Delta_{\mathrm{FD}}$. The maximum absolute errors in the total $S_0$
gradient appear to converge roughly geometrically in both the size of the
finite difference stencil $n_{\mathrm{FD}}$ and the inverse of the finite
difference spacing $\Delta_{\mathrm{FD}}^{-1}$. Remarkably course finite
difference grids with, e.g., $(n_{\mathrm{FD}}, \Delta_{FD}) = (4, 0.2)$ already
exhibit maximum absolute deviations in the gradient of $<10^{-6}$, i.e., far
smaller than the magnitude of the response terms. 

\begin{figure}[ht!]
\begin{center}
\includegraphics[width=3.4in]{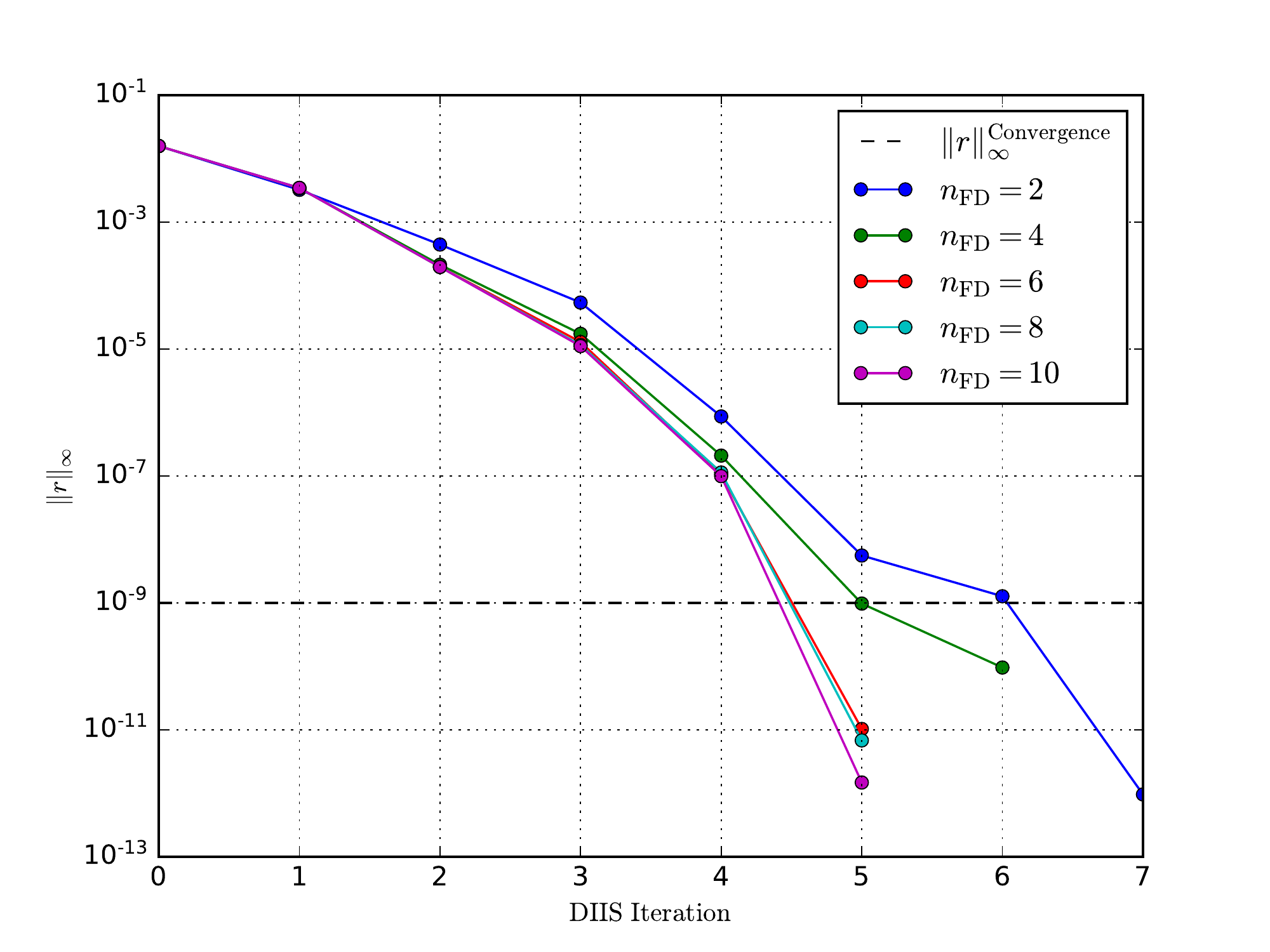}
\caption{Convergence in the SA-VQE response equations for $S_0$ state of
cyclohexadiene ring-closing MECI computed with FOMO-CAS-MC-VQE(6e, 4o)/6-31G*
with 1 complete layer in the QNP entangler circuit. Here the SA-VQE response
residual is shown as a function of DIIS iteration for various finite difference
stencil sizes $n_{\mathrm{FD}}$ for finite different step size of
$\Delta_{\mathrm{FD}} = 0.3$. THe DIIS equations terminate at a convergence
criterion of $| \vec r |_{\infty} < 10^{-9}$.}
\label{fig:diis_convergence-1}
\end{center}
\end{figure}

\begin{figure}[ht!]
\begin{center}
\includegraphics[width=3.4in]{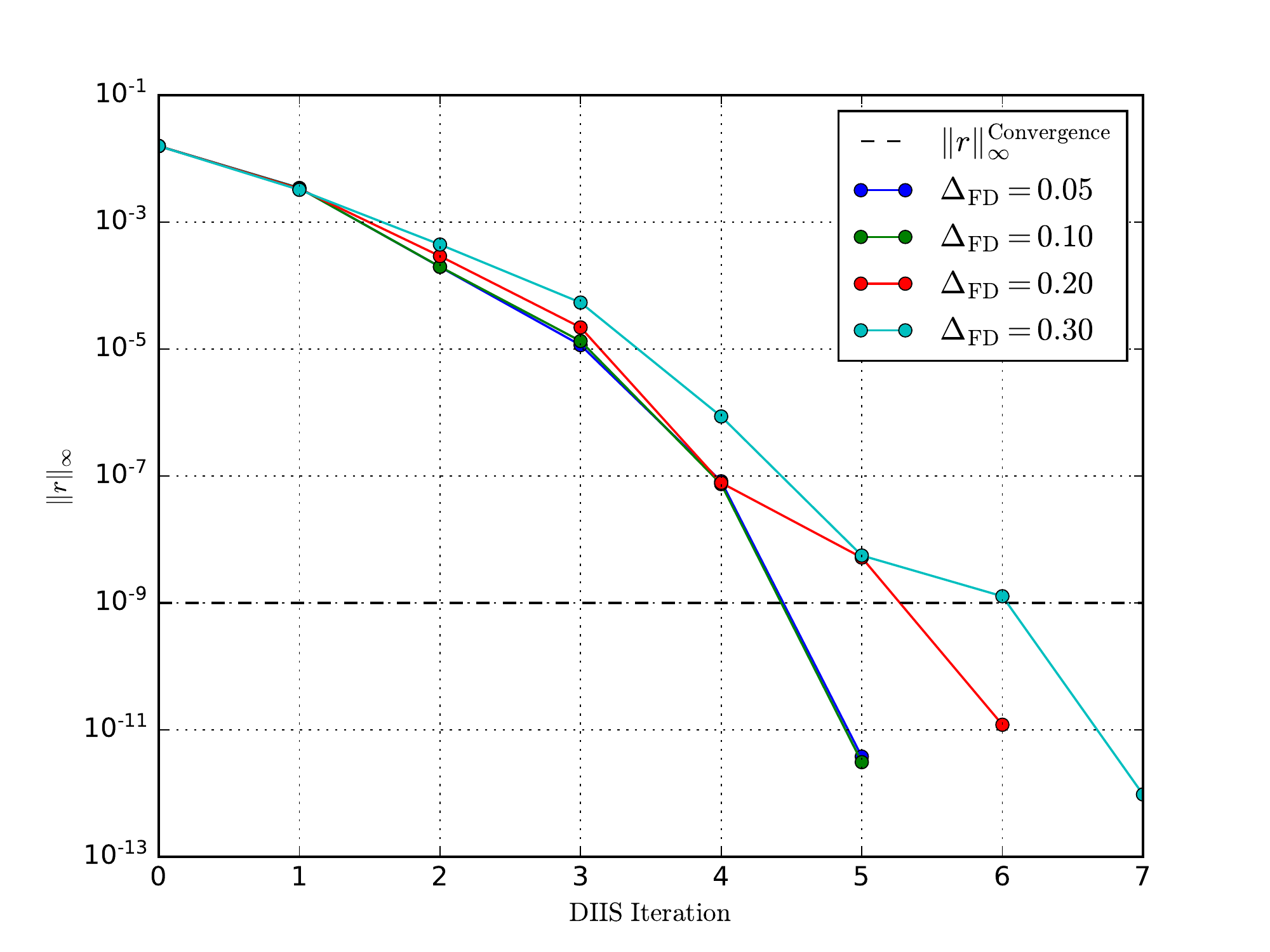}
\caption{Convergence in the SA-VQE response equations for $S_0$ state of
cyclohexadiene ring-closing MECI computed with FOMO-CAS-MC-VQE(6e, 4o)/6-31G*
with 1 complete layer in the QNP entangler circuit. Here the SA-VQE response
residual is shown as a function of DIIS iteration for various finite difference
step sizes $\Delta_{\mathrm{FD}}$ for finite difference stencil size of
$n_{\mathrm{FD}} = 2$. THe DIIS equations terminate at a convergence
criterion of $| \vec r |_{\infty} < 10^{-9}$.}
\label{fig:diis_convergence-2}
\end{center}
\end{figure}

One additional potential concern is that the extrinsic introduction of finite
difference matrix-vector product formalism might significantly inhibit the
convergence of the DIIS iterative procedure. This is addressed in Figures
\ref{fig:diis_convergence-1} and \ref{fig:diis_convergence-2}. Here it is found
that medium-quality finite difference stencils all show the same convergence
behavior, which is qualitative indistinguishable from that of ideal
matrix-vector products. Only the coarsest finite difference grids in
$\Delta_{\mathrm{FD}}$ and $n_{\mathrm{FD}}$ exhibit noticeable deviations from
the ideal convergence behavior. These require only 1 to 2 additional iterations
to converge, and the early convergence behavior that this most relevant for the
shot-noise limited deployment of VQE is quite similar for all cases.

\subsection{Large-Scale Test Case}

For a large-scale test case we consider octatetraene solvated by 50 MeOH solvent
molecules, with the geometry from \cite{snyder2017direct}, with 318 atoms. We
compute the orbitals with FON-RHF($\beta=0.15$, Gaussian broadening)/6-31G,
which comprises 1392 atomic orbitals. This level of theory for classical orbital
determination and active space integral preparation/derivative postprocessing
is tractable within a few minutes of GPU-accelerated compute time. An active
space of (6e, 6o) orbitals is used - note that the intuitively-preferred active
space of (8e, 8o) orbitals experiences orbital instabilities between the LUMO+4
and LUMO+5 orbitals with this basis set and geometry.

\subsection{Large-Scale Results}

\begin{figure}[ht!]
\begin{center}
\includegraphics[width=\linewidth]{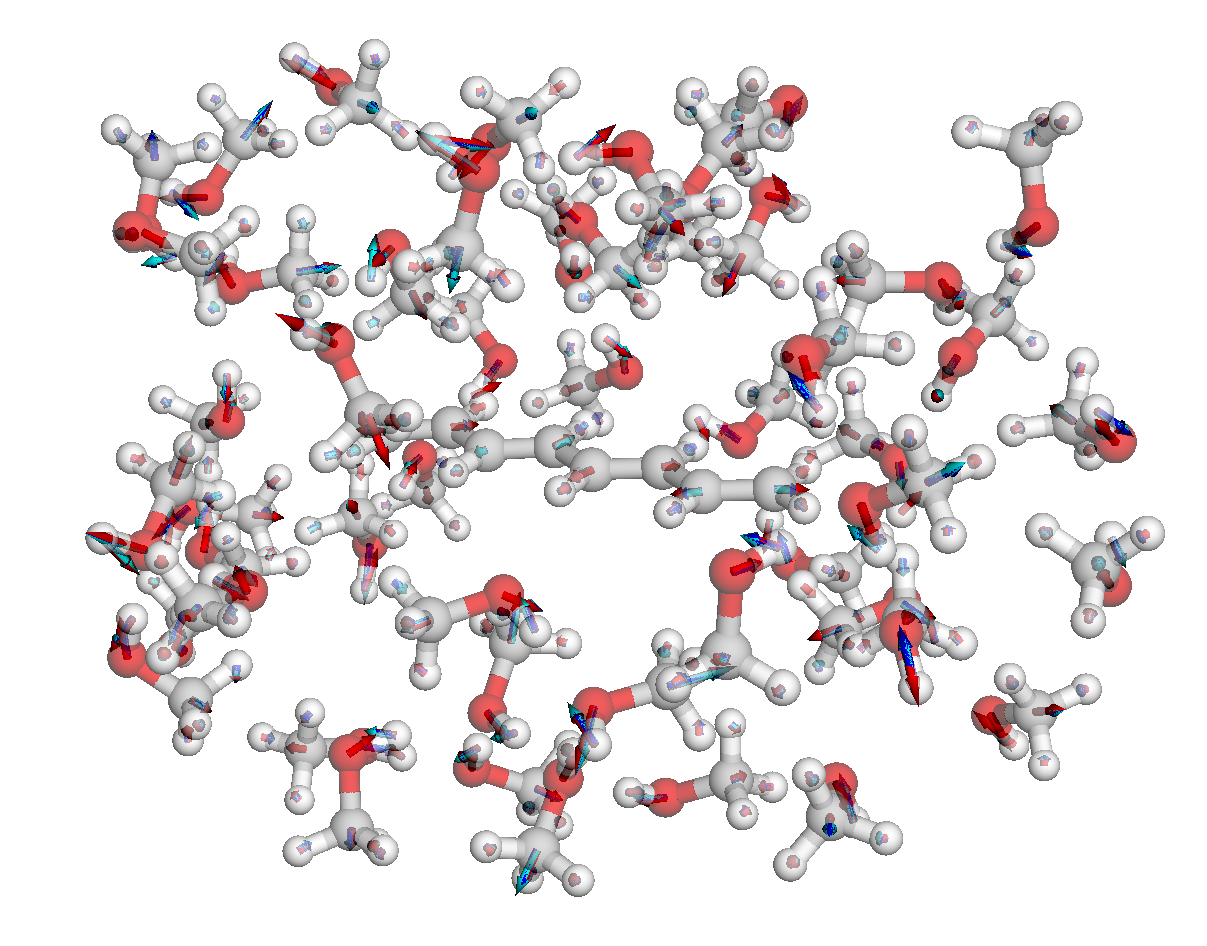}
\includegraphics[width=\linewidth]{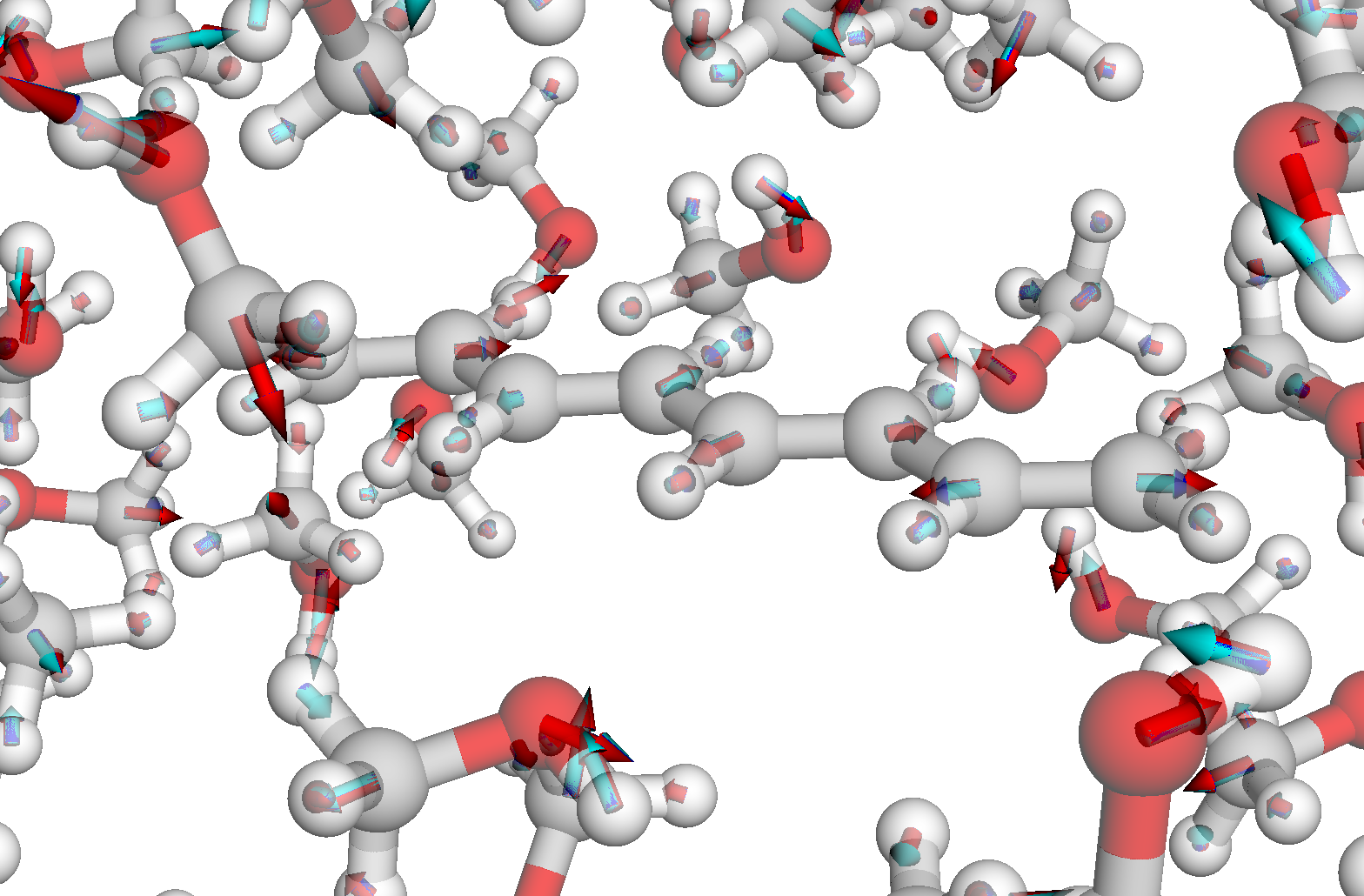}
\caption{Gradients of singlet ground state of (6e, 6o) active space of
octatetraene@MeOH$_{50}$ with orbitals computed at FON-RHF($\beta=0.15$,
Gaussian broadening)/6-31G.  Within the active space, gradients are computed
with FCI (red), MC-VQE including SA-VQE response (blue) and MC-VQE without
SA-VQE response (turquoise). Three total layers of QNP entangler gates are used
in SA-VQE response. (Upper panel) Full system. (Lower panel) Inset zoomed in on
octatetraene.}
\label{fig:oct50}
\end{center}
\end{figure}

Figure \ref{fig:oct50} depicts the ground-state gradients of the
octatetraene@MeOH$_50$ test case with a (6e, 6o) active space. Particularly, we
compare FCI treatment of the active space wavefunction (red), MC-VQE with two
states averaged in SA-VQE and three total layers of QNP entangler gates (blue),
and the same MC-VQE without response terms (turquoise). The full Hessian is
used in the computation of the response terms, i.e., no finite difference
approximations are used. Generally, all three layers of theory agree
qualitatively, with some exceptions on the nearest OH group in the prominent
lower right MeOH solvent molecule (where an FON-RHF active orbital is
hybridizing with those on the octatetraene). The agreement between
SA-vQE-response-including and bare MC-VQE gradients are generally within $\sim
2\%$ of the maximum gradient element, and this agreement is much tighter than
the agreement between CASCI and MC-VQE for this short depth of SA-VQE entangler.
The main conclusion is that external classical system size is not a major issue
in computing analytical MC-VQE gradients. 

\section{Summary and Outlook}

In the present work, we have laid out a complete procedure for the efficient
evaluation of analytical first-order derivative properties for nonstationary
VQE-type methods with active space embeddings. The use of the Lagrangian
formalism is found to remove formal scaling dependencies on the number of
derivative perturbations in the computation of the response terms, and also
maximally separates the quantum and classical response contributions. 
Importantly it makes the quantum computational effort independent of the
total number of derivative properties, which, e.g., enables the computation of
analytical gradients of systems of hundreds of atoms, whose few most important 
orbitals are accurately simulated on a quantum computer. 

The use of a novel iterative SA-VQE Hessian matrix-vector product formalism with
widely-spaced finite difference approximations for the needed matrix-vector
products is found to remove the need for explicit treatment of the SA-VQE
Hessian while retaining an accurate solution of the SA-VQE response equations.
Taken together, the Lagrangian approach and iterative matrix-vector product
formalism appear to yield a method with similar computational resource
requirements for the energy and gradient. Such methodology can be relatively
easily adapted to other hybrid quantum/classical approaches, including other
non-stationary algorithms such as certain variants of QSE-VQE, SS-VQE, NO-VQE,
QFD, or VQPE, and can be extended to other underlying classical methods such as
CIS-NO \cite{shu2015configuration,fales2017complete} or UNO
\cite{bofill1989unrestricted} orbital determination. Additionally many other
target observables and derivative perturbations can be efficiently accessed
with this methodology, including polarizabilities, non-adiabatic coupling
vectors, and NMR chemical shifts.

There are many interesting directions to pursue from this point. One is an
obvious need for a thorough quantification and mitigation of shot-noise and
decoherence-noise errors in all steps of the energy and gradient computation. It
may well be that much more efficient procedures can be developed if
consideration of quantum circuit shot cost is woven throughout the methodology,
as has been explored by a number of authors in other context. e.g., Bayesian
methods.  Another straightforward direction to consider is the use of
more-advanced representations of the qubit-basis Hamiltonian operator to reduce
measurement burden. Double factorization is surely an interesting direction to
pursue along these lines,
though additional classical response terms may arise
to reflect the approximate nature of the truncated, tensor factorized
Hamiltonian in such cases. Higher derivatives could be considered, though it
should be noted that in classical electronic structure methodology, higher
derivatives are much less prevalent than first derivatives due to the verbosity
of implementation and necessary increase in computational scaling when pursuing
second and higher-order derivatives.

Another important question is where
exactly the SA-VQE response terms need to be explicitly computed. For small
systems with deep SA-VQE entangler circuits on very near-term hardware, the
shot noise and decoherence noise channels will surely dominate the error, and
the simple expedient of using unrelaxed VQE density matrices can be sufficient.
I.e., in the simple cases considered herein, the response terms constituted
only $\sim 2\%$ of the total gradient.  In the longer term, in our opinion
(based on numerous experiences with classical methods where response terms can
become quite significant in larger systems), it is likely that response terms
in VQE methods will eventually become required.

\textbf{Acknowledgements:} QC Ware Corp. acknowledges generous research funding
from Covestro Deutschland AG for this project.  Covestro acknowledges funding
from the German Ministry for Education and Research (BMBF) under the funding
program quantum technologies as part of project HFAK (13N15630). 

\textbf{Conflict of Interest:} The hybrid quantum/classical Lagrangian
technique described in this work is the focus of a US provisional application
for patent filed jointly by QC Ware Corp. and Covestro Deutschland AG. RMP and
Covestro Deutschland AG own stock/options in QC Ware Corp.

\bibliographystyle{unsrturl}
\bibliography{jrncodes,cg,gian,rmp,rmp2,cg2,yo}

\begin{thebibliography}{10}

\bibitem{Helgaker:1982:LAG}
Trygve~Ulf Helgaker.
\newblock Simple derivation of the potential energy gradient for an arbitrary
  electronic wave function.
\newblock {\em Int. J. Quant. Chem.}, 21(5):939--940, 1982.
\newblock \href {http://dx.doi.org/10.1002/qua.560210520}
  {\path{doi:10.1002/qua.560210520}}.

\bibitem{Dalgarno:1958:245}
Alexander Dalgarno and AL~Stewart.
\newblock A perturbation calculation of properties of the helium iso-electronic
  sequence.
\newblock {\em Proceedings of the Royal Society of London. Series A.
  Mathematical and Physical Sciences}, 247(1249):245--259, 1958.
\newblock \href {http://dx.doi.org/10.1098/rspa.1958.0182}
  {\path{doi:10.1098/rspa.1958.0182}}.

\bibitem{Handy:1984:Zvec}
Nicholas~C Handy and Henry~F Schaefer~III.
\newblock On the evaluation of analytic energy derivatives for correlated wave
  functions.
\newblock {\em J. Chem. Phys.}, 81(11):5031--5033, 1984.
\newblock \href {http://dx.doi.org/10.1063/1.447489}
  {\path{doi:10.1063/1.447489}}.

\bibitem{Schaefer:1986:new}
Henry~F Schaefer~III and Yukio Yamaguchi.
\newblock A new dimension to quantum chemistry: Theoretical methods for the
  analytic evaluation of first, second, and third derivatives of the molecular
  electronic energy with respect to nuclear coordinates.
\newblock {\em J. Mol. Struct.}, 135:369--390, 1986.
\newblock \href {http://dx.doi.org/10.1021/ed072pA42.6}
  {\path{doi:10.1021/ed072pA42.6}}.

\bibitem{Kassal:2009:grad}
Ivan Kassal and Al{\'a}n Aspuru-Guzik.
\newblock Quantum algorithm for molecular properties and geometry optimization.
\newblock {\em J. Chem. Phys.}, 131(22):224102, 2009.
\newblock \href {http://dx.doi.org/10.1063/1.3266959}
  {\path{doi:10.1063/1.3266959}}.

\bibitem{preskill2018quantum}
John Preskill.
\newblock Quantum computing in the nisq era and beyond.
\newblock {\em Quantum}, 2:79, 2018.
\newblock URL: \url{https://doi.org/10.22331/q-2018-08-06-79}, \href
  {http://dx.doi.org/10.22331/q-2018-08-06-79}
  {\path{doi:10.22331/q-2018-08-06-79}}.

\bibitem{mcclean2017hybrid}
Jarrod~R McClean, Mollie~E Kimchi-Schwartz, Jonathan Carter, and Wibe~A
  De~Jong.
\newblock Hybrid quantum-classical hierarchy for mitigation of decoherence and
  determination of excited states.
\newblock {\em Physical Review A}, 95(4):042308, 2017.
\newblock URL: \url{https://doi.org/10.1103/PhysRevA.95.042308}, \href
  {http://dx.doi.org/10.1103/PhysRevA.95.042308}
  {\path{doi:10.1103/PhysRevA.95.042308}}.

\bibitem{parrish2019quantum}
Robert~M Parrish, Edward~G Hohenstein, Peter~L McMahon, and Todd~J
  Mart{\'\i}nez.
\newblock Quantum computation of electronic transitions using a variational
  quantum eigensolver.
\newblock {\em Physical review letters}, 122(23):230401, 2019.
\newblock URL: \url{https://doi.org/10.1103/PhysRevLett.122.230401}, \href
  {http://dx.doi.org/10.1103/PhysRevLett.122.230401}
  {\path{doi:10.1103/PhysRevLett.122.230401}}.

\bibitem{nakanishi2019subspace}
Ken~M Nakanishi, Kosuke Mitarai, and Keisuke Fujii.
\newblock Subspace-search variational quantum eigensolver for excited states.
\newblock {\em Physical Review Research}, 1(3):033062, 2019.
\newblock URL: \url{https://doi.org/10.1103/PhysRevResearch.1.033062}, \href
  {http://dx.doi.org/10.1103/PhysRevResearch.1.033062}
  {\path{doi:10.1103/PhysRevResearch.1.033062}}.

\bibitem{parrish2019qfd}
Robert~M Parrish and Peter~L McMahon.
\newblock Quantum filter diagonalization: Quantum eigendecomposition without
  full quantum phase estimation.
\newblock {\em arXiv preprint arXiv:1909.08925}, 2019.
\newblock URL: \url{https://arxiv.org/abs/1909.08925}.

\bibitem{urbanek2020chemistry}
Miroslav Urbanek, Daan Camps, Roel Van~Beeumen, and Wibe~A de~Jong.
\newblock Chemistry on quantum computers with virtual quantum subspace
  expansion.
\newblock {\em Journal of Chemical Theory and Computation}, 16(9):5425--5431,
  2020.
\newblock URL: \url{https://doi.org/10.1021/acs.jctc.0c00447}, \href
  {http://dx.doi.org/10.1021/acs.jctc.0c00447}
  {\path{doi:10.1021/acs.jctc.0c00447}}.

\bibitem{ollitrault2020quantum}
Pauline~J Ollitrault, Abhinav Kandala, Chun-Fu Chen, Panagiotis~Kl Barkoutsos,
  Antonio Mezzacapo, Marco Pistoia, Sarah Sheldon, Stefan Woerner, Jay~M
  Gambetta, and Ivano Tavernelli.
\newblock Quantum equation of motion for computing molecular excitation
  energies on a noisy quantum processor.
\newblock {\em Physical Review Research}, 2(4):043140, 2020.
\newblock URL: \url{https://doi.org/10.1103/PhysRevResearch.2.043140}, \href
  {http://dx.doi.org/10.1103/PhysRevResearch.2.043140}
  {\path{doi:10.1103/PhysRevResearch.2.043140}}.

\bibitem{huggins2020non}
William~J Huggins, Joonho Lee, Unpil Baek, Bryan O’Gorman, and K~Birgitta
  Whaley.
\newblock A non-orthogonal variational quantum eigensolver.
\newblock {\em New Journal of Physics}, 22(7):073009, 2020.
\newblock \href {http://dx.doi.org/10.1088/1367-2630/ab867b}
  {\path{doi:10.1088/1367-2630/ab867b}}.

\bibitem{stair2020multireference}
Nicholas~H Stair, Renke Huang, and Francesco~A Evangelista.
\newblock A multireference quantum krylov algorithm for strongly correlated
  electrons.
\newblock {\em Journal of chemical theory and computation}, 16(4):2236--2245,
  2020.
\newblock URL: \url{https://doi.org/10.1021/acs.jctc.9b01125}, \href
  {http://dx.doi.org/10.1021/acs.jctc.9b01125}
  {\path{doi:10.1021/acs.jctc.9b01125}}.

\bibitem{klymko2021real}
Katherine Klymko, Carlos Mejuto-Zaera, Stephen~J Cotton, Filip Wudarski,
  Miroslav Urbanek, Diptarka Hait, Martin Head-Gordon, K~Birgitta Whaley,
  Jonathan Moussa, Nathan Wiebe, et~al.
\newblock Real time evolution for ultracompact hamiltonian eigenstates on
  quantum hardware.
\newblock {\em arXiv preprint arXiv:2103.08563}, 2021.
\newblock \href {http://arxiv.org/abs/https://arxiv.org/abs/2103.08563}
  {\path{arXiv:https://arxiv.org/abs/2103.08563}}.

\bibitem{McClean:2016:023023}
Jarrod~R McClean, Jonathan Romero, Ryan Babbush, and Al{\'a}n Aspuru-Guzik.
\newblock The theory of variational hybrid quantum-classical algorithms.
\newblock {\em New J. Phys.}, 18(2):023023, 2016.
\newblock \href {http://dx.doi.org/10.1088/1367-2630/18/2/023023}
  {\path{doi:10.1088/1367-2630/18/2/023023}}.

\bibitem{delgado2021variational}
Alain Delgado, Juan~Miguel Arrazola, Soran Jahangiri, Zeyue Niu, Josh Izaac,
  Chase Roberts, and Nathan Killoran.
\newblock Variational quantum algorithm for molecular geometry optimization,
  2021.
\newblock \href {http://arxiv.org/abs/2106.13840} {\path{arXiv:2106.13840}}.

\bibitem{mitarai2020theory}
Kosuke Mitarai, Yuya~O Nakagawa, and Wataru Mizukami.
\newblock Theory of analytical energy derivatives for the variational quantum
  eigensolver.
\newblock {\em Physical Review Research}, 2(1):013129, 2020.
\newblock \href {http://dx.doi.org/10.1103/PhysRevResearch.2.013129}
  {\path{doi:10.1103/PhysRevResearch.2.013129}}.

\bibitem{Azad2021}
Utkarsh Azad and Harjinder Singh.
\newblock Quantum chemistry calculations using energy derivatives on quantum
  computers.
\newblock {\em arXiv preprint arXiv:2106.06463}, 2021.
\newblock \href {http://arxiv.org/abs/2106.06463} {\path{arXiv:2106.06463}}.

\bibitem{o2019calculating}
Thomas~E O’Brien, Bruno Senjean, Ramiro Sagastizabal, Xavier Bonet-Monroig,
  Alicja Dutkiewicz, Francesco Buda, Leonardo DiCarlo, and Lucas Visscher.
\newblock Calculating energy derivatives for quantum chemistry on a quantum
  computer.
\newblock {\em npj Quantum Information}, 5(1):1--12, 2019.
\newblock \href {http://dx.doi.org/10.1038/s41534-019-0213-4}
  {\path{doi:10.1038/s41534-019-0213-4}}.

\bibitem{parrish2019hybrid}
Robert~M. Parrish, Edward~G. Hohenstein, Peter~L. McMahon, and Todd~J.
  Martinez.
\newblock Hybrid quantum/classical derivative theory: Analytical gradients and
  excited-state dynamics for the multistate contracted variational quantum
  eigensolver, 2019.
\newblock \href {http://arxiv.org/abs/1906.08728} {\path{arXiv:1906.08728}}.

\bibitem{arimitsu2021analytic}
Keita Arimitsu, Yuya~O Nakagawa, Sho Koh, Wataru Mizukami, Qi~Gao, and Takao
  Kobayashi.
\newblock Analytic energy gradient for state-averaged orbital-optimized
  variational quantum eigensolvers and its application to a photochemical
  reaction.
\newblock {\em arXiv preprint arXiv:2107.12705}, 2021.
\newblock \href {http://arxiv.org/abs/https://arxiv.org/abs/2107.12705}
  {\path{arXiv:https://arxiv.org/abs/2107.12705}}.

\bibitem{yalouz2021analytical}
Saad Yalouz, Emiel Koridon, Bruno Senjean, Benjamin Lasorne, Francesco Buda,
  and Lucas Visscher.
\newblock Analytical nonadiabatic couplings and gradients within the
  state-averaged orbital-optimized variational quantum eigensolver.
\newblock {\em arXiv preprint arXiv:2109.04576}, 2021.
\newblock \href {http://arxiv.org/abs/https://arxiv.org/abs/2109.04576}
  {\path{arXiv:https://arxiv.org/abs/2109.04576}}.

\bibitem{gidopoulos1994hartree}
Nikitas Gidopoulos and Andreas Theophilou.
\newblock Hartree-fock equations determining the optimum set of spin orbitals
  for the expansion of excited states.
\newblock {\em Philosophical Magazine B}, 69(5):1067--1074, 1994.
\newblock \href {http://dx.doi.org/10.1080/01418639408240176}
  {\path{doi:10.1080/01418639408240176}}.

\bibitem{gidopoulos2002ensemble}
Nikitas~I Gidopoulos, Petros~G Papaconstantinou, and Eberhardt~KU Gross.
\newblock Ensemble-hartree--fock scheme for excited states. the optimized
  effective potential method.
\newblock {\em Physica B: Condensed Matter}, 318(4):328--332, 2002.
\newblock \href {http://dx.doi.org/10.1016/S0921-4526(02)00799-8}
  {\path{doi:10.1016/S0921-4526(02)00799-8}}.

\bibitem{barbatti2006ultrafast}
Mario Barbatti, Ad{\'e}lia~JA Aquino, and Hans Lischka.
\newblock Ultrafast two-step process in the non-adiabatic relaxation of the ch2
  molecule.
\newblock {\em Molecular Physics}, 104(5-7):1053--1060, 2006.
\newblock \href {http://dx.doi.org/10.1080/00268970500417945}
  {\path{doi:10.1080/00268970500417945}}.

\bibitem{sellner2013ultrafast}
Bernhard Sellner, Mario Barbatti, Thomas M{\"u}ller, Wolfgang Domcke, and Hans
  Lischka.
\newblock Ultrafast non-adiabatic dynamics of ethylene including rydberg
  states.
\newblock {\em Molecular physics}, 111(16-17):2439--2450, 2013.
\newblock \href {http://dx.doi.org/10.1080/00268976.2013.813590}
  {\path{doi:10.1080/00268976.2013.813590}}.

\bibitem{hohenstein2015analytic}
Edward~G Hohenstein, Marine~EF Bouduban, Chenchen Song, Nathan Luehr, Ivan~S
  Ufimtsev, and Todd~J Mart{\'\i}nez.
\newblock Analytic first derivatives of floating occupation molecular
  orbital-complete active space configuration interaction on graphical
  processing units.
\newblock {\em The Journal of chemical physics}, 143(1):014111, 2015.
\newblock \href {http://dx.doi.org/10.1063/1.4923259}
  {\path{doi:10.1063/1.4923259}}.

\bibitem{hohenstein2016analytic}
Edward~G Hohenstein.
\newblock Analytic formulation of derivative coupling vectors for complete
  active space configuration interaction wavefunctions with floating occupation
  molecular orbitals.
\newblock {\em The Journal of chemical physics}, 145(17):174110, 2016.
\newblock \href {http://dx.doi.org/10.1063/1.4966235}
  {\path{doi:10.1063/1.4966235}}.

\bibitem{QNP2021}
Gian-Luca~R. Anselmetti, David Wierichs, Christian Gogolin, and Robert~M.
  Parrish.
\newblock Local, expressive, quantum-number-preserving vqe ansatze for
  fermionic systems.
\newblock {\em arXiv preprint arXiv:2104.05695}, 2021.
\newblock \href {http://dx.doi.org/10.1088/1367-2630/ac2cb3}
  {\path{doi:10.1088/1367-2630/ac2cb3}}.

\bibitem{peruzzo2014variational}
Alberto Peruzzo, Jarrod McClean, Peter Shadbolt, Man-Hong Yung, Xiao-Qi Zhou,
  Peter~J Love, Al{\'a}n Aspuru-Guzik, and Jeremy~L O’brien.
\newblock A variational eigenvalue solver on a photonic quantum processor.
\newblock {\em Nature communications}, 5(1):1--7, 2014.
\newblock \href {http://dx.doi.org/10.1038/ncomms5213}
  {\path{doi:10.1038/ncomms5213}}.

\bibitem{PhysRevX.6.031007}
P.~J.~J. O'Malley, R.~Babbush, I.~D. Kivlichan, J.~Romero, J.~R. McClean,
  R.~Barends, J.~Kelly, P.~Roushan, A.~Tranter, N.~Ding, B.~Campbell, Y.~Chen,
  Z.~Chen, B.~Chiaro, A.~Dunsworth, A.~G. Fowler, E.~Jeffrey, E.~Lucero,
  A.~Megrant, J.~Y. Mutus, M.~Neeley, C.~Neill, C.~Quintana, D.~Sank,
  A.~Vainsencher, J.~Wenner, T.~C. White, P.~V. Coveney, P.~J. Love, H.~Neven,
  A.~Aspuru-Guzik, and J.~M. Martinis.
\newblock Scalable quantum simulation of molecular energies.
\newblock {\em Phys. Rev. X}, 6:031007, Jul 2016.
\newblock URL: \url{https://link.aps.org/doi/10.1103/PhysRevX.6.031007}, \href
  {http://dx.doi.org/10.1103/PhysRevX.6.031007}
  {\path{doi:10.1103/PhysRevX.6.031007}}.

\bibitem{Kandala2017}
Abhinav Kandala, Antonio Mezzacapo, Kristan Temme, Maika Takita, Markus Brink,
  Jerry~M. Chow, and Jay~M. Gambetta.
\newblock Hardware-efficient variational quantum eigensolver for small
  molecules and quantum magnets.
\newblock {\em Nature}, 549(7671):242--246, September 2017.
\newblock \href {http://dx.doi.org/10.1038/nature23879}
  {\path{doi:10.1038/nature23879}}.

\bibitem{Ryabinkin2018}
Ilya~G. Ryabinkin, Tzu-Ching Yen, Scott~N. Genin, and Artur~F. Izmaylov.
\newblock Qubit coupled cluster method: A systematic approach to quantum
  chemistry on a quantum computer.
\newblock {\em Journal of Chemical Theory and Computation}, 14(12):6317--6326,
  November 2018.
\newblock URL: \url{https://doi.org/10.1021/acs.jctc.8b00932}, \href
  {http://dx.doi.org/10.1021/acs.jctc.8b00932}
  {\path{doi:10.1021/acs.jctc.8b00932}}.

\bibitem{lee2018generalized}
Joonho Lee, William~J Huggins, Martin Head-Gordon, and K~Birgitta Whaley.
\newblock Generalized unitary coupled cluster wave functions for quantum
  computation.
\newblock {\em Journal of chemical theory and computation}, 15(1):311--324,
  2018.
\newblock URL: \url{https://pubs.acs.org/doi/abs/10.1021/acs.jctc.8b01004}.

\bibitem{evangelista2019exact}
Francesco~A Evangelista, Garnet Kin-Lic Chan, and Gustavo~E Scuseria.
\newblock Exact parameterization of fermionic wave functions via unitary
  coupled cluster theory.
\newblock {\em The Journal of chemical physics}, 151(24):244112, 2019.
\newblock URL: \url{https://doi.org/10.1063/1.5133059}, \href
  {http://dx.doi.org/10.1063/1.5133059} {\path{doi:10.1063/1.5133059}}.

\bibitem{ganzhorn2019gate}
Marc Ganzhorn, Daniel~J Egger, Panagiotis Barkoutsos, Pauline Ollitrault, Gian
  Salis, Nikolaj Moll, M~Roth, A~Fuhrer, P~Mueller, Stefan Woerner, et~al.
\newblock Gate-efficient simulation of molecular eigenstates on a quantum
  computer.
\newblock {\em Physical Review Applied}, 11(4):044092, 2019.
\newblock URL: \url{https://doi.org/10.1103/PhysRevApplied.11.044092}, \href
  {http://dx.doi.org/10.1103/PhysRevApplied.11.044092}
  {\path{doi:10.1103/PhysRevApplied.11.044092}}.

\bibitem{salis2019short}
Gian Salis and Nikolaj Moll.
\newblock Short-depth trial-wavefunctions for the variational quantum
  eigensolver based on the problem hamiltonian.
\newblock {\em arXiv preprint arXiv:1908.09533}, 2019.
\newblock URL: \url{https://arxiv.org/abs/1908.09533}.

\bibitem{Bian2019}
Teng Bian, Daniel Murphy, Rongxin Xia, Ammar Daskin, and Sabre Kais.
\newblock Quantum computing methods for electronic states of the water
  molecule.
\newblock {\em Molecular Physics}, 117(15-16):2069--2082, February 2019.
\newblock URL: \url{https://doi.org/10.1080/00268976.2019.1580392}, \href
  {http://dx.doi.org/10.1080/00268976.2019.1580392}
  {\path{doi:10.1080/00268976.2019.1580392}}.

\bibitem{o2019generalized}
Bryan O'Gorman, William~J Huggins, Eleanor~G Rieffel, and K~Birgitta Whaley.
\newblock Generalized swap networks for near-term quantum computing.
\newblock {\em arXiv preprint arXiv:1905.05118}, 2019.
\newblock URL: \url{https://arxiv.org/abs/1905.05118}.

\bibitem{gard2020efficient}
Bryan~T Gard, Linghua Zhu, George~S Barron, Nicholas~J Mayhall, Sophia~E
  Economou, and Edwin Barnes.
\newblock Efficient symmetry-preserving state preparation circuits for the
  variational quantum eigensolver algorithm.
\newblock {\em npj Quantum Information}, 6(1):1--9, 2020.
\newblock \href {http://dx.doi.org/10.1038/s41534-019-0240-1}
  {\path{doi:10.1038/s41534-019-0240-1}}.

\bibitem{xia2020qubit}
Rongxin Xia and Sabre Kais.
\newblock Qubit coupled cluster singles and doubles variational quantum
  eigensolver ansatz for electronic structure calculations.
\newblock {\em Quantum Science and Technology}, 6(1):015001, 2020.
\newblock \href {http://dx.doi.org/10.1088/2058-9565/abbc74}
  {\path{doi:10.1088/2058-9565/abbc74}}.

\bibitem{yordanov2020efficient}
Yordan~S Yordanov, David~RM Arvidsson-Shukur, and Crispin~HW Barnes.
\newblock Efficient quantum circuits for quantum computational chemistry.
\newblock {\em Physical Review A}, 102(6):062612, 2020.
\newblock URL: \url{https://doi.org/10.1103/PhysRevA.102.062612}, \href
  {http://dx.doi.org/10.1103/PhysRevA.102.062612}
  {\path{doi:10.1103/PhysRevA.102.062612}}.

\bibitem{khamoshi2020correlating}
Armin Khamoshi, Francesco~A Evangelista, and Gustavo~E Scuseria.
\newblock Correlating agp on a quantum computer.
\newblock {\em Quantum Science and Technology}, 6(1):014004, 2020.
\newblock URL: \url{https://doi.org/10.1088/2058-9565/abc1bb}, \href
  {http://dx.doi.org/10.1088/2058-9565/abc1bb}
  {\path{doi:10.1088/2058-9565/abc1bb}}.

\bibitem{matsuzawa2020jastrow}
Yuta Matsuzawa and Yuki Kurashige.
\newblock Jastrow-type decomposition in quantum chemistry for low-depth quantum
  circuits.
\newblock {\em Journal of chemical theory and computation}, 16(2):944--952,
  2020.
\newblock URL: \url{https://doi.org/10.1021/acs.jctc.9b00963}, \href
  {http://dx.doi.org/10.1021/acs.jctc.9b00963}
  {\path{doi:10.1021/acs.jctc.9b00963}}.

\bibitem{poulin2014trotter}
David Poulin, Matthew~B. Hastings, Dave Wecker, Nathan Wiebe, Andrew~C.
  Doberty, and Matthias Troyer.
\newblock The trotter step size required for accurate quantum simulation of
  quantum chemistry.
\newblock 15(5–6):361–384, April 2015.
\newblock URL: \url{https://dl.acm.org/doi/10.5555/2871401.2871402}.

\bibitem{peng2017highly}
Bo~Peng and Karol Kowalski.
\newblock Highly efficient and scalable compound decomposition of two-electron
  integral tensor and its application in coupled cluster calculations.
\newblock 13(9):4179--4192, 2017.
\newblock URL: \url{https://pubs.acs.org/doi/10.1021/acs.jctc.7b00605}.

\bibitem{motta2018low}
Mario Motta, Erika Ye, Jarrod~R. McClean, Zhendong Li, Austin~J. Minnich, Ryan
  Babbush, and Garnet Kin-Lic Chan.
\newblock Low rank representations for quantum simulation of electronic
  structure.
\newblock {\em npj Quantum Information}, 7(1):83, May 2021.
\newblock \href {http://dx.doi.org/10.1038/s41534-021-00416-z}
  {\path{doi:10.1038/s41534-021-00416-z}}.

\bibitem{motta2019efficient}
Mario Motta, James Shee, Shiwei Zhang, and Garnet Kin-Lic Chan.
\newblock Efficient ab initio auxiliary-field quantum monte carlo calculations
  in gaussian bases via low-rank tensor decomposition.
\newblock 15(6):3510--3521, 2019.
\newblock URL: \url{https://pubs.acs.org/doi/abs/10.1021/acs.jctc.8b00996}.

\bibitem{kivlichan2018quantum}
Ian~D Kivlichan, Jarrod McClean, Nathan Wiebe, Craig Gidney, Al{\'a}n
  Aspuru-Guzik, Garnet Kin-Lic Chan, and Ryan Babbush.
\newblock Quantum simulation of electronic structure with linear depth and
  connectivity.
\newblock 120(11):110501, 2018.
\newblock URL:
  \url{https://journals.aps.org/prl/abstract/10.1103/PhysRevLett.120.110501}.

\bibitem{berry2019qubitization}
Dominic~W Berry, Craig Gidney, Mario Motta, Jarrod~R McClean, and Ryan Babbush.
\newblock Qubitization of arbitrary basis quantum chemistry leveraging sparsity
  and low rank factorization.
\newblock {\em Quantum}, 3:208, 2019.
\newblock URL: \url{https://quantum-journal.org/papers/q-2019-12-02-208/}.

\bibitem{huggins2021efficient}
William~J Huggins, Jarrod~R McClean, Nicholas~C Rubin, Zhang Jiang, Nathan
  Wiebe, K~Birgitta Whaley, and Ryan Babbush.
\newblock Efficient and noise resilient measurements for quantum chemistry on
  near-term quantum computers.
\newblock 7(1):1--9, 2021.
\newblock URL: \url{https://www.nature.com/articles/s41534-020-00341-7}.

\bibitem{cohn2021quantum}
Jeffrey Cohn, Mario Motta, and Robert~M Parrish.
\newblock Quantum filter diagonalization with double-factorized hamiltonians.
\newblock {\em arXiv preprint arXiv:2104.08957}, 2021.
\newblock \href {http://arxiv.org/abs/https://arxiv.org/abs/2104.08957}
  {\path{arXiv:https://arxiv.org/abs/2104.08957}}.

\bibitem{PhysRevLett.118.150503}
Jun Li, Xiaodong Yang, Xinhua Peng, and Chang-Pu Sun.
\newblock Hybrid quantum-classical approach to quantum optimal control.
\newblock {\em Phys. Rev. Lett.}, 118:150503, Apr 2017.
\newblock URL: \url{https://link.aps.org/doi/10.1103/PhysRevLett.118.150503},
  \href {http://dx.doi.org/10.1103/PhysRevLett.118.150503}
  {\path{doi:10.1103/PhysRevLett.118.150503}}.

\bibitem{PhysRevA.98.032309}
K.~Mitarai, M.~Negoro, M.~Kitagawa, and K.~Fujii.
\newblock Quantum circuit learning.
\newblock {\em Phys. Rev. A}, 98:032309, Sep 2018.
\newblock URL: \url{https://link.aps.org/doi/10.1103/PhysRevA.98.032309}, \href
  {http://dx.doi.org/10.1103/PhysRevA.98.032309}
  {\path{doi:10.1103/PhysRevA.98.032309}}.

\bibitem{PennyLane}
Ville Bergholm, Josh Izaac, Maria Schuld, Christian Gogolin, M~Sohaib Alam,
  Shahnawaz Ahmed, Juan~Miguel Arrazola, Carsten Blank, Alain Delgado, Soran
  Jahangiri, et~al.
\newblock Pennylane: Automatic differentiation of hybrid quantum-classical
  computations.
\newblock {\em arXiv preprint arXiv:1811.04968}, 2018.
\newblock URL: \url{https://arxiv.org/abs/1811.04968}.

\bibitem{Mari2021}
Andrea Mari, Thomas~R. Bromley, and Nathan Killoran.
\newblock Estimating the gradient and higher-order derivatives on quantum
  hardware.
\newblock {\em Physical Review A}, 103(1), January 2021.
\newblock URL: \url{https://doi.org/10.1103/physreva.103.012405}, \href
  {http://dx.doi.org/10.1103/physreva.103.012405}
  {\path{doi:10.1103/physreva.103.012405}}.

\bibitem{wierichs2021general}
David Wierichs, Josh Izaac, Cody Wang, and Cedric Yen-Yu Lin.
\newblock General parameter-shift rules for quantum gradients, 2021.
\newblock \href {http://arxiv.org/abs/2107.12390} {\path{arXiv:2107.12390}}.

\bibitem{Hubregtsen2021}
Thomas Hubregtsen, Frederik Wilde, Shozab Qasim, and Jens Eisert.
\newblock Single-component gradient rules for variational quantum algorithms.
\newblock {\em arXiv preprint arXiv:2106.01388}, 2021.
\newblock \href {http://arxiv.org/abs/2106.01388} {\path{arXiv:2106.01388}}.

\bibitem{Pulay:1980:393}
Péter Pulay.
\newblock Convergence acceleration of iterative sequences. the case of scf
  iteration.
\newblock {\em Chem. Phys. Lett.}, 73(2):393 -- 398, 1980.
\newblock \href
  {http://dx.doi.org/https://doi.org/10.1016/0009-2614(80)80396-4}
  {\path{doi:https://doi.org/10.1016/0009-2614(80)80396-4}}.

\bibitem{Pulay:1982:556}
P.~Pulay.
\newblock Improved scf convergence acceleration.
\newblock {\em J. Comp. Chem.}, 3(4):556--560, 1982.
\newblock \href {http://dx.doi.org/10.1002/jcc.540030413}
  {\path{doi:10.1002/jcc.540030413}}.

\bibitem{csaszar1984geometry}
P{\'a}l Cs{\'a}sz{\'a}r and P{\'e}ter Pulay.
\newblock Geometry optimization by direct inversion in the iterative subspace.
\newblock {\em J. Mol. Struct.}, 114:31--34, 1984.
\newblock \href {http://dx.doi.org/10.1016/S0022-2860(84)87198-7}
  {\path{doi:10.1016/S0022-2860(84)87198-7}}.

\bibitem{hamilton1986direct}
Tracy~P Hamilton and Peter Pulay.
\newblock Direct inversion in the iterative subspace (diis) optimization of
  open-shell, excited-state, and small multiconfiguration scf wave functions.
\newblock {\em J. Chem. Phys.}, 84(10):5728--5734, 1986.
\newblock \href {http://dx.doi.org/10.1063/1.449880}
  {\path{doi:10.1063/1.449880}}.

\bibitem{hutter1994electronic}
J{\"u}rg Hutter, Hans~Peter L{\"u}thi, and Michele Parrinello.
\newblock Electronic structure optimization in plane-wave-based density
  functional calculations by direct inversion in the iterative subspace.
\newblock {\em Comp. Mat. Sci.}, 2(2):244--248, 1994.
\newblock \href {http://dx.doi.org/10.1016/0927-0256(94)90105-8}
  {\path{doi:10.1016/0927-0256(94)90105-8}}.

\bibitem{scuseria1986accelerating}
Gustavo~E Scuseria, Timothy~J Lee, and Henry~F Schaefer~III.
\newblock Accelerating the convergence of the coupled-cluster approach: The use
  of the diis method.
\newblock {\em Chem. Phys. Lett.}, 130(3):236--239, 1986.
\newblock \href {http://dx.doi.org/10.1016/0009-2614(86)80461-4}
  {\path{doi:10.1016/0009-2614(86)80461-4}}.

\bibitem{kudin2002black}
Konstantin~N Kudin, Gustavo~E Scuseria, and Eric Cances.
\newblock A black-box self-consistent field convergence algorithm: One step
  closer.
\newblock {\em J. Chem. Phys.}, 116(19):8255--8261, 2002.
\newblock \href {http://dx.doi.org/10.1063/1.1470195}
  {\path{doi:10.1063/1.1470195}}.

\bibitem{chen2011listb}
Ya~Kun Chen and Yan~Alexander Wang.
\newblock Listb: a better direct approach to list.
\newblock {\em J. Chem. Theory Comput.}, 7(10):3045--3048, 2011.
\newblock \href {http://dx.doi.org/10.1021/ct2004512}
  {\path{doi:10.1021/ct2004512}}.

\bibitem{hu2017projected}
Wei Hu, Lin Lin, and Chao Yang.
\newblock Projected commutator diis method for accelerating hybrid functional
  electronic structure calculations.
\newblock {\em J. Chem. Theory Comput.}, 13(11):5458--5467, 2017.
\newblock \href {http://dx.doi.org/10.1021/acs.jctc.7b00892}
  {\path{doi:10.1021/acs.jctc.7b00892}}.

\bibitem{parrish2019jacobi}
Robert~M Parrish, Joseph~T Iosue, Asier Ozaeta, and Peter~L McMahon.
\newblock A jacobi diagonalization and anderson acceleration algorithm for
  variational quantum algorithm parameter optimization.
\newblock {\em arXiv preprint arXiv:1904.03206}, 2019.
\newblock \href {http://arxiv.org/abs/https://arxiv.org/abs/1904.03206}
  {\path{arXiv:https://arxiv.org/abs/1904.03206}}.

\bibitem{rubin2021fermionic}
Nicholas~C Rubin, Toru Shiozaki, Kyle Throssell, Garnet~Kin Chan, and Ryan
  Babbush.
\newblock The fermionic quantum emulator.
\newblock {\em arXiv preprint arXiv:2104.13944}, 2021.
\newblock \href {http://arxiv.org/abs/https://arxiv.org/abs/2104.1394}
  {\path{arXiv:https://arxiv.org/abs/2104.1394}}.

\bibitem{stair2021qforte}
Nicholas~H Stair and Francesco~A Evangelista.
\newblock Qforte: an efficient state simulator and quantum algorithms library
  for molecular electronic structure.
\newblock {\em arXiv preprint arXiv:2108.04413}, 2021.
\newblock \href {http://arxiv.org/abs/https://arxiv.org/abs/2108.04413}
  {\path{arXiv:https://arxiv.org/abs/2108.04413}}.

\bibitem{OpenFermion}
Jarrod~R McClean, Nicholas~C Rubin, Kevin~J Sung, Ian~D Kivlichan, Xavier
  Bonet-Monroig, Yudong Cao, Chengyu Dai, E~Schuyler Fried, Craig Gidney,
  Brendan Gimby, Pranav Gokhale, Thomas H\"{a}ner, Tarini Hardikar,
  Vojt{\v{e}}ch Havl{\'{\i}}{\v{c}}ek, Oscar Higgott, Cupjin Huang, Josh Izaac,
  Zhang Jiang, Xinle Liu, Sam McArdle, Matthew Neeley, Thomas O'Brien, Bryan
  O'Gorman, Isil Ozfidan, Maxwell~D Radin, Jhonathan Romero, Nicolas P~D
  Sawaya, Bruno Senjean, Kanav Setia, Sukin Sim, Damian~S Steiger, Mark
  Steudtner, Qiming Sun, Wei Sun, Daochen Wang, Fang Zhang, and Ryan Babbush.
\newblock {OpenFermion}: the electronic structure package for quantum
  computers.
\newblock {\em Quantum Science and Technology}, 5(3):034014, June 2020.
\newblock URL: \url{https://doi.org/10.1088/2058-9565/ab8ebc}, \href
  {http://dx.doi.org/10.1088/2058-9565/ab8ebc}
  {\path{doi:10.1088/2058-9565/ab8ebc}}.

\bibitem{wolf2019photochemical}
Thomas~JA Wolf, David~M Sanchez, J~Yang, RM~Parrish, JPF Nunes, M~Centurion,
  R~Coffee, JP~Cryan, Markus G{\"u}hr, Kareem Hegazy, et~al.
\newblock The photochemical ring-opening of 1, 3-cyclohexadiene imaged by
  ultrafast electron diffraction.
\newblock {\em Nature chemistry}, 11(6):504--509, 2019.
\newblock \href {http://dx.doi.org/10.1038/s41557-019-0252-7}
  {\path{doi:10.1038/s41557-019-0252-7}}.

\bibitem{snyder2017direct}
James~W Snyder~Jr, B~Scott Fales, Edward~G Hohenstein, Benjamin~G Levine, and
  Todd~J Mart{\'\i}nez.
\newblock A direct-compatible formulation of the coupled perturbed complete
  active space self-consistent field equations on graphical processing units.
\newblock {\em The Journal of chemical physics}, 146(17):174113, 2017.

\bibitem{shu2015configuration}
Yinan Shu, Edward~G Hohenstein, and Benjamin~G Levine.
\newblock Configuration interaction singles natural orbitals: An orbital basis
  for an efficient and size intensive multireference description of electronic
  excited states.
\newblock {\em The Journal of chemical physics}, 142(2):024102, 2015.
\newblock \href {http://dx.doi.org/10.1063/1.4905124}
  {\path{doi:10.1063/1.4905124}}.

\bibitem{fales2017complete}
B~Scott Fales, Yinan Shu, Benjamin~G Levine, and Edward~G Hohenstein.
\newblock Complete active space configuration interaction from state-averaged
  configuration interaction singles natural orbitals: Analytic first
  derivatives and derivative coupling vectors.
\newblock {\em The Journal of chemical physics}, 147(9):094104, 2017.
\newblock \href {http://dx.doi.org/10.1063/1.5000476}
  {\path{doi:10.1063/1.5000476}}.

\bibitem{bofill1989unrestricted}
Josep~M Bofill and Peter Pulay.
\newblock The unrestricted natural orbital--complete active space (uno--cas)
  method: An inexpensive alternative to the complete active
  space--self-consistent-field (cas--scf) method.
\newblock {\em The Journal of Chemical Physics}, 90(7):3637--3646, 1989.
\newblock \href {http://dx.doi.org/10.1063/1.455822}
  {\path{doi:10.1063/1.455822}}.

\end{thebibliography}

\newpage
\clearpage

\appendix

\section{Technical Notes: Specific FON-RHF-MC-VQE Ansatz Choices}

\subsection{Orbital Determination: Fractional Occupation Number Restricted
Hartree Fock (FON-RHF)}

\label{sec:FON-RHF}

The fractional occupation number restricted Hartree Fock (FON-RHF) method
constructs the real spatial molecular orbitals $\{ \phi_{p} (\vec r_1) \equiv
\sum_{\mu} C_{\mu p} \chi_{\mu} (\vec r_1) \}$ from the real non-orthogonal
atomic orbitals $\{ \chi_{\mu} (\vec r_1) \}$, subject to constraints to
diagonalize the Fock matrix,
\begin{equation}
f_{pq}
\equiv
(p|\hat h|q)
+
\sum_{r}
2 (pq|rr) n_{r}
-
\sum_{r}
(pr|qr) n_{r}
=
\delta_{pq}
\epsilon_{p}
\end{equation}
to keep the orbitals orthonormal,
\begin{equation}
(p|q)
= 
\delta_{pq}
\end{equation}
and to preserve a target total number of occupied electrons,
\begin{equation}
\sum_{r}
n_{r}
=
N_{\mathrm{FOMO}}
\end{equation}
Note that $N_{\mathrm{FOMO}}$ is a user-specified parameter that in practice may
be different than the total number of electrons in a subsequent CASCI procedure.
However, herein, we use the simple choice of $N_{\mathrm{FOMO}} \equiv
N_{\mathrm{T}}$.

The fractional orbital occupation number $n_{r} \in [0, 1]$ for each orbital
index $r$ is determined by a strictly nonincreasing function in the orbital
energy $\epsilon_{r}$. Several popular occupation number functions are used,
including the Fermi-Dirac cutoff function,
\begin{equation}
n_{r}^{\mathrm{Fermi-Dirac}}
\equiv
\frac{1}{
1 + \exp[\beta (\epsilon_{r} - \mu) ]
}
\end{equation}
and the Gaussian smearing cutoff function,
\begin{equation}
n_{r}^{\mathrm{Gaussian}}
\equiv
\frac{1}{2}
\erfc
\left [
\beta (\epsilon_{r} - \mu)
\right ]
\end{equation}
In both of these cases, the constant $\beta$ is supplied by the user, and may be
roughly interpreted as inverse electronic temperature. The wavefunction
parameter $\mu$ is varied to conserve the total electron number constraint, and
may be roughly interpreted as the Fermi energy. The FON-RHF occupation numbers
may be applied within an active space picture, i.e., the user may elect to clamp
the occupation numbers to $1$ for a set of core orbitals, the occupation numbers
to $0$ for a set of virtual orbitals, and to let the smooth cutoff procedure
provide fractional occupation numbers for a set of active orbitals. Note that
the partitions into core/active/virtual subsets may be different from those used
in any subsequent CASCI operations, though herein we use the simple choice of
coincident core/active/virtual partitions for both the FON-RHF and CASCI/MC-VQE
portions of the method.

The FON-RHF orbitals are fully analytically differentiable (except at a
topologically small subset of parameter cases involved edge core/active and
active/virtual degeneracies in occupation numbers), as shown in several works by
other authors. Typically a CASCI analytical gradient procedure built on FON-RHF
orbitals involves the following steps: (1) determination of the relaxed density
matrix in the active space (e.g., including SA-VQE response) (2) determination
of the diagonal relaxed CASCI OPDM and TPDM contributions to the gradient (3)
construction and solution of the CP-FON-RHF equations (i.e., classical orbital
response) (4) accumulation of the orbital response contributions to the gradient
through modified Fock matrix gradients. In particular, once the active-space
relaxed MC-VQE TPDM is determined (relaxed to include SA-VQE response), the
existing classical FOMO-CASCI gradient code stack handles all of the rest of the
computation of the gradient, including all details of the chain-rule gradient
contributions of the FON-RHF orbital response, spatial molecular integrals, and
nuclear positions. The interested reader is referred for example to Hohenstein's
FOMO-CASCI gradient paper for more details \cite{hohenstein2015analytic}.

\subsection{External System Embedding: Restricted Hartree-Fock}

\label{sec:EMB-RHF}

Once the orbitals $\{ \phi_{p} (\vec r_1) \}$ have been determined, we partition
them into core ($i$, $j$, $k$, $l$), active ($p$, $q$, $r$, $s$) and virtual
($a$, $b$, $c$, $d$) subsets. Using conventional RHF-level embedding of the core
orbitals, the active space molecular integrals are,
\begin{equation}
E_{\mathrm{ext}}
\equiv
E_{\mathrm{Nuc}}
+
\sum_{i}
2 (i|\hat h|i)
+
\sum_{ij}
2 (ii|jj)
-
\sum_{ij}
(ij|ij)
\end{equation}
and the one-body Hamiltonian is,
\begin{equation}
(p|\hat h|q)
\equiv
(p|- \nabla_1^2 / 2 + v_{\mathrm{Nuc}} (\vec r_1)|q)
+
2 
\sum_{i}
(pq|ii)
-
\sum_{i}
(pi|qi)
\end{equation}
Here $E_{\mathrm{Nuc}}$ is the nuclear-nuclear self interaction energy, and
$v_{\mathrm{Nuc}} (\vec r_1)$ is the electrostatic potential of the nuclei.

\subsection{MC-VQE Reference State Determination: Configuration State Functions}

With the spatial orbitals defined and partitioned, we now need to define a
pragmatic set of active-space reference states for MC-VQE. Desired
characteristics of these states include:
\begin{itemize}
\item Classical tractability: These states must be efficiently able to be
determined and represented by a polynomial number of classical operations. 
\item Quantum tractability: These states should be preparable by simple/short quantum
circuits, with all needed quantum circuit gate parameters determined \emph{a
priori} by classical computation.
\item Orthogonality: For the specific flavor of MC-VQE used here, we require
that the reference states be orthonormal. This restriction could be lifted in
future work by using a metric-based variant of MC-VQE.
\item Physical Relevance: Insofar as possible, these states should approximately
span the space of the target Hamiltonian eigenfunctions.
\item Quantum number preservation: These states must all be proper
eigenfunctions of $\hat N_{\alpha}$, $\hat N_{\beta}$, and $\hat S^2$ with the
target quantum numbers as eigenvalues.
\end{itemize}

One pragmatic (but certainly not unique) choice for MC-VQE in fermionic systems
is the set of certain configuration state functions (CSFs) in the given qubit
orbital basis (i.e., the FON-RHF orbital basis). For instance, for singlet
states, we will include the RHF CSF,
\begin{equation}
| \Phi_{0} \rangle
\equiv
\sum_{i}^{N_{\alpha}}
i^\dagger
\bar i^\dagger
| \rangle
\end{equation}
the set of singlet singly-excited CSFs,
\begin{equation}
| \Phi_{i \rightarrow a}^{\mathrm{S}} \rangle
\equiv
\frac{1}{\sqrt{2}}
\left [
a^\dagger i
+
\bar a^\dagger \bar i
\right ]
| \Phi_{0} \rangle
\end{equation}
and the set of diagonal doubly-excited CSFs,
\begin{equation}
| \Phi_{i \rightarrow a}^{\mathrm{D}} \rangle
\equiv
a^\dagger i \bar a^\dagger \bar i
| \Phi_{0} \rangle
\end{equation}
For a given target number of MC-VQE states, we limit ourselves to a discrete
subset of these via a selection procedure. For instance, we may elect to sort
the energies of the reference states and then take lowest few sorted reference
states to be the working set for MC-VQE. We may also choose the reference states
according to character, i.e., by choosing those that maximize the overlap with
states from a nearby geometry during dynamics or geometry optimization.

\section{Technical Notes: Jordan Wigner Hamiltonian}

\label{sec:JW}

For complete details on the logically $\alpha$-then-$\beta$ and physically
interleaved flavor of the Jordan-Wigner mapping used herein, and for additional
definitions of the Jordan-Wigner forms of the fermionic composition,
substitution, $\hat N_{\alpha}$, $\hat N_{\beta}$, and $\hat S^2$ operators,
the reader is referred to Appendix B of our manuscript on quantum number
preserving entangler circuit.

In order to provide a completely closed procedure for the computation of the
active space density matrices, we must provide a differentiable mapping between
the second-quantized Hamiltonian operator and corresponding qubit compatible
operators such as Pauli operators. The primary reason for this is to provide a
closed recipe for the backtransformation from qubit-basis observables to the
unrelaxed OPDM/TPDM.  E.g., if the Hamiltonian is written in Pauli words $\{
\hat \Pi_{I} \}$ as $\hat H \equiv \sum_{I} \mathcal{H}_{I} \hat \Pi_{I}$, and
we define the Pauli-basis unrelaxed density matrix elements $\{ \Gamma_{I} \}$
as,
\begin{equation}
\Gamma_{I}
\equiv
\langle \Psi^{\Theta} | \hat \Pi_{I} | \Psi^{\Theta} \rangle
\end{equation}
then we can exploit the equality of the energy trace formulae,
\begin{equation}
E^{\Theta}
=
E_{\mathrm{ext}}
+
\sum_{pq}
\gamma_{pq}^{\Theta}
(p | \hat h | q)
+
\frac{1}{2}
\sum_{pqrs}
\Gamma_{pqrs}^{\Theta}
(pq|rs)
\end{equation}
\[
=
\sum_{I}
\mathcal{H}_{I}
\Gamma_{I}
\]
to obtain the backtransformation formulae,
\begin{equation}
\Rightarrow
\gamma_{pq}^{\Theta}
=
\sum_{I}
\pdiff{
\mathcal{H}_{I}
}{(p|\hat h|q)}
\Gamma_{I}^{\Theta}
\end{equation}
and,
\begin{equation}
\Rightarrow
\Gamma_{pqrs}^{\Theta}
=
2
\sum_{I}
\pdiff{
\mathcal{H}_{I}
}{(pq|rs)}
\Gamma_{I}^{\Theta}
\end{equation}
These backtransformation formulae are highly sparse, and are easily performed in
classical postprocessing.  Below, we show the simplified definition of the
Hamiltonian operator in terms of qubit Pauli operators under the Jordan-Wigner
mapping. 

As discussed in the main text, such a laborious enumeration in Pauli operators
is conceptually straightforward, but has been superseded in numerical practice
by more-advanced techniques such as double factorization (DF). This does not
present any theoretical barrier to computing the active space density matrices,
so long as the qubit basis operators of the more advanced representation are
fully differentiable with respect to $(p|\hat h|q)$ and $(pq|rs)$. 

After expansion in labelled spin orbitals, the Hamiltonian is,
\begin{equation}
\hat H
=
\sum_{pq}
(p|\hat h|q)
\left (
p^{+}
q
+
\bar p^{+}
\bar q
\right )
\end{equation}
\[
+ 
\frac{1}{2}
\sum_{pqrs}
(pq|rs)
\left [
p^+
r^+
s
q
+
p^+
\bar r^+
\bar s
q
+
\bar p^+
r^+
s
\bar q
+
\bar p^+
\bar r^+
\bar s
\bar q
\right ]
\]
\[
=
\sum_{pq}
(p|\hat h|q)
\left (
p^{+}
q
+
\bar p^{+}
\bar q
\right )
\]
\[
+ 
\frac{1}{2}
\sum_{pqrs}
(pq|rs)
\left [
2
\bar p^+
\bar q
r^+
s
+
p^+
r^+
s
q
+
\bar p^+
\bar r^+
\bar s
\bar q
\right ]
\]

The following terms enter the Jordan-Wigner Hamiltonian:

One-particle $\alpha$,
\begin{equation}
\hat H 
\leftarrow
+
\frac{1}{2}
\sum_{p}
(p|\hat h|p)
\hat I
-
\frac{1}{2}
\sum_{p}
(p|\hat h|p)
\hat Z_{p}
\end{equation}
\[
+
\frac{1}{2}
\sum_{p<q}
(p|\hat h|q)
\hat X_{p}
\otimes
\hat Z_{p+1,q-1}^{\leftrightarrow}
\otimes
\hat X_{q}
\]
\[
+
\frac{1}{2}
\sum_{p<q}
(p|\hat h|q)
\hat Y_{p}
\otimes
\hat Z_{p+1,q-1}^{\leftrightarrow}
\otimes
\hat Y_{q}
\]
One-particle $\beta$,
\begin{equation}
\hat H 
\leftarrow
+
\frac{1}{2}
\sum_{p}
(p|\hat h|p)
\hat I
-
\frac{1}{2}
\sum_{p}
(p|\hat h|p)
\hat Z_{\bar p}
\end{equation}
\[
+
\frac{1}{2}
\sum_{p<q}
(p|\hat h|q)
\hat X_{\bar p}
\otimes
\hat Z_{\bar p+1,\bar q-1}^{\leftrightarrow}
\otimes
\hat X_{\bar q}
\]
\[
+
\frac{1}{2}
\sum_{p<q}
(p|\hat h|q)
\hat Y_{\bar p}
\otimes
\hat Z_{\bar p+1,\bar q-1}^{\leftrightarrow}
\otimes
\hat Y_{\bar q}
\]

Two-particle $\alpha$-$\beta$,

\begin{equation}
\hat H
\leftarrow
\frac{1}{4}
\sum_{p}
\sum_{r}
(pp|rr)
-
\frac{1}{4}
\sum_{p}
\sum_{r}
(pp|rr)
\hat Z_{\bar r}
\end{equation}
\[
-
\frac{1}{4}
\sum_{p}
\sum_{r}
(pp|rr)
\hat Z_{p}
+
\frac{1}{4}
\sum_{p}
\sum_{r}
(pp|rr)
\hat Z_{p}
\otimes
\hat Z_{\bar r}
\]
\[
+
\frac{1}{4}
\sum_{p}
\sum_{r < s}
(pp|rs)
\hat X_{\bar r}
\leftrightarrow
\hat X_{\bar s}
+
\frac{1}{4}
\sum_{p}
\sum_{r < s}
(pp|rs)
\hat Y_{\bar r}
\leftrightarrow
\hat Y_{\bar s}
\]
\[
-
\frac{1}{4}
\sum_{p}
\sum_{r < s}
(pp|rs)
\hat Z_{p}
\hat X_{\bar r}
\leftrightarrow
\hat X_{\bar s}
-
\frac{1}{4}
\sum_{p}
\sum_{r < s}
(pp|rs)
\hat Z_{p}
\hat Y_{\bar r}
\leftrightarrow
\hat Y_{\bar s}
\]
\[
+
\frac{1}{4}
\sum_{p < q}
\sum_{r}
(pq|rr)
\hat X_{p}
\leftrightarrow
\hat X_{q}
+
\frac{1}{4}
\sum_{p < q}
\sum_{r}
(pq|rr)
\hat Y_{p}
\leftrightarrow
\hat Y_{q}
\]
\[
-
\frac{1}{4}
\sum_{p < q}
\sum_{r}
(pq|rr)
\hat Z_{\bar r}
\hat X_{p}
\leftrightarrow
\hat X_{q}
-
\frac{1}{4}
\sum_{p < q}
\sum_{r}
(pq|rr)
\hat Z_{\bar r}
\hat Y_{p}
\leftrightarrow
\hat Y_{q}
\]
\[
+
\frac{1}{4}
\sum_{p < q}
\sum_{r < s}
(pq|rs)
\hat X_{p}
\leftrightarrow
\hat X_{q}
\hat X_{\bar r}
\leftrightarrow
\hat X_{\bar s}
\]
\[
+
\frac{1}{4}
\sum_{p < q}
\sum_{r < s}
(pq|rs)
\hat X_{p}
\leftrightarrow
\hat X_{q}
\hat Y_{\bar r}
\leftrightarrow
\hat Y_{\bar s}
\]
\[
+
\frac{1}{4}
\sum_{p < q}
\sum_{r < s}
(pq|rs)
\hat Y_{p}
\leftrightarrow
\hat Y_{q}
\hat X_{\bar r}
\leftrightarrow
\hat X_{\bar s}
\]
\[
+
\frac{1}{4}
\sum_{p < q}
\sum_{r < s}
(pq|rs)
\hat Y_{p}
\leftrightarrow
\hat Y_{q}
\hat Y_{\bar r}
\leftrightarrow
\hat Y_{\bar s}
\]

The same-spin terms are, e.g., for $\alpha\alpha$,
\begin{equation}
\hat H
\leftarrow
\frac{1}{4}
\sum_{p < q}
\langle pq || pq \rangle
\end{equation}
\[
-
\frac{1}{4}
\sum_{p}
\left [
\sum_{q}
\langle pq || pq \rangle
\right ]
\hat Z_{p}
\]
\[
+
\frac{1}{4}
\sum_{p < q}
\langle pq || pq \rangle
\hat Z_{p} 
\otimes 
\hat Z_{q}
\]
\[
\frac{1}{4}
\sum_{q < r}
\left [
\sum_{p}
\langle pq || pr \rangle
\right ]
\hat X_{q} \leftrightarrow \hat X_{r}
\]
\[
+
\frac{1}{4}
\sum_{q < r}
\left [
\sum_{p}
\langle pq || pr \rangle
\right ]
\hat Y_{q} \leftrightarrow \hat Y_{r}
\]
\[
-
\frac{1}{4}
\sum_{q < r}
\sum_{p \neq q, r}
\langle pq || pr \rangle
\hat Z_{p}
\hat X_{q} \leftrightarrow \hat X_{r}
\]
\[
-
\frac{1}{4}
\sum_{q < r}
\sum_{p \neq q, r}
\langle pq || pr \rangle
\hat Z_{p}
\hat Y_{q} \leftrightarrow \hat Y_{r}
\]
\[
% 1
+
\frac{1}{4}
\sum_{p < q < r < s}
[
(pq|rs) 
- 
(ps|qr)
]
X_{p} \leftrightarrow X_{q} X_{r} \leftrightarrow X_{s}
\]
% 2
\[
+
\frac{1}{4}
\sum_{p < q < r < s}
[
(pq|rs) 
- 
(pr|qs)
]
X_{p} \leftrightarrow X_{q} Y_{r} \leftrightarrow Y_{s}
\]
% 5
\[
+
\frac{1}{4}
\sum_{p < q < r < s}
[
(pr|qs) 
- 
(ps|qr)
]
X_{p} \leftrightarrow Y_{q} Y_{r} \leftrightarrow X_{s}
\]
% 6
\[
+
\frac{1}{4}
\sum_{p < q < r < s}
[
(pr|qs) 
- 
(ps|qr)
]
Y_{p} \leftrightarrow X_{q} X_{r} \leftrightarrow Y_{s}
\]
% 3
\[
+
\frac{1}{4}
\sum_{p < q < r < s}
[
(pq|rs) 
- 
(pr|qs)
]
Y_{p} \leftrightarrow Y_{q} X_{r} \leftrightarrow X_{s}
\]
% 4
\[
+
\frac{1}{4}
\sum_{p < q < r < s}
[
(pq|rs) 
- 
(ps|qr)
]
Y_{p} \leftrightarrow Y_{q} Y_{r} \leftrightarrow Y_{s}
\]
And similarly in $\beta \beta$ by ``barring'' of indices.

Here, the antisymmetrized two-electron integrals are,
\begin{equation}
\langle pq || rs \rangle
\equiv
\langle pq | rs \rangle
-
\langle pq | sr \rangle
=
(pr|qs)
-
(ps|qr)
\end{equation}
Note the 8-fold antisymmetry,
\begin{equation}
\langle pq || rs \rangle
=
-
\langle pq || sr \rangle
=
-
\langle qp || rs \rangle
=
\langle qp || sr \rangle
\end{equation}
\[
=
\langle rs || pq \rangle
=
-
\langle rs || qp \rangle
=
-
\langle sr || pq \rangle
=
\langle sr || qp \rangle
\]
Also note that the antisymmetry mandates,
\begin{equation}
\langle pp || rs \rangle
=
\langle pq || rr \rangle
=
0
\end{equation}

\subsection{Detailed Derivation of Same-Spin Two-Electron Hamiltonian Term}

The same-spin two-electron component of the Hamiltonian can be written (e.g., in
$\alpha$) as,
\begin{equation}
\hat H^{\alpha \alpha}
\equiv
\sum_{pqrs}
\frac{1}{4}
\langle pq || rs \rangle
p^+
q^+
s 
r
\end{equation}
\[
=
\sum_{p < q}
\sum_{r < s}
\langle pq || rs \rangle
p^+
q^+
s 
r
\]

First up, the cases with full contraction,
$(p = r),\  (q = s)$, and
$(p = s),\  (q = r)$, 
but (enforced by the antisymmetrized integrals) $(p \neq q)$ and $(r \neq s)$.
\begin{equation}
H^{\alpha \alpha}
\leftarrow
\frac{1}{4}
\sum_{pq}
\langle pq || pq \rangle
p^+ 
q^+
q
p
+
\langle pq || qp \rangle
p^+ 
q^+
p
q
\end{equation}
\[
=
\frac{1}{2}
\sum_{pq}
\langle pq || pq \rangle
p^+ 
p
q^+
q
\]
\[
=
\frac{1}{8}
\sum_{pq}
\langle pq || pq \rangle
( \hat I_{p} - \hat Z_{p} )
( \hat I_{q} - \hat Z_{q} )
\]
\[
=
\frac{1}{4}
\sum_{p < q}
\langle pq || pq \rangle
-
\frac{1}{4}
\sum_{p}
\left (
\sum_{q}
\langle pq || pq \rangle
\right )
\hat Z_{p}
\]
\[
+
\frac{1}{4}
\sum_{p < q}
\langle pq || pq \rangle
\hat Z_{p} 
\otimes 
\hat Z_{q}
\]

Now, let us consider the case with a single contraction. There are four cases:
\\
$p = r$ $p \neq q$, $r \neq s$ $\Rightarrow p \neq s$ \\
$p = s$ $p \neq q$, $r \neq s$ $\Rightarrow p \neq r$ \\
$q = r$ $p \neq q$, $r \neq s$ $\Rightarrow q \neq s$ \\ 
$q = s$ $p \neq q$, $r \neq s$ $\Rightarrow q \neq r$ \\

Below $\sum_{pqr}^{\prime} \equiv \sum_{p \neq q \neq r}$,
\begin{equation}
H^{\alpha \alpha}
\leftarrow
\frac{1}{4}
\sum_{pqs}^{\prime}
\langle pq || ps \rangle
p^+ 
q^+
s 
p
+
\frac{1}{4}
\sum_{pqr}^{\prime}
\langle pq || rp \rangle
p^+ 
q^+
p 
r
\end{equation}
\[
+
\frac{1}{4}
\sum_{pqs}^{\prime}
\langle pq || qs \rangle
p^+ 
q^+
s 
q
+
\frac{1}{4}
\sum_{pqr}^{\prime}
\langle pq || rq \rangle
p^+ 
q^+
q 
r
\]
\[
=
\sum_{pqr}^{\prime}
\langle pq || pr \rangle
p^+ 
p
q^+
r 
\]
\[
=
\frac{1}{4}
\sum_{p}
\sum_{q < r}
\langle pq || pr \rangle
(
\hat I_p 
-
\hat Z_p
)
\left [
\hat X_{q}
\leftrightarrow
\hat X_{r}
+
\hat Y_{q}
\leftrightarrow
\hat Y_{r}
\right ]
\]
\[
=
\frac{1}{4}
\sum_{q < r}
\left [
\sum_{p}
\langle pq || pr \rangle
\right ]
\hat X_{q} \leftrightarrow \hat X_{r}
+
\frac{1}{4}
\sum_{q < r}
\left [
\sum_{p}
\langle pq || pr \rangle
\right ]
\hat Y_{q} \leftrightarrow \hat Y_{r}
\]
\[
-
\frac{1}{4}
\sum_{q < r}
\sum_{p \neq q, r}
\langle pq || pr \rangle
\hat Z_{p}
\hat X_{q} \leftrightarrow \hat X_{r}
-
\frac{1}{4}
\sum_{q < r}
\sum_{p \neq q, r}
\langle pq || pr \rangle
\hat Z_{p}
\hat Y_{q} \leftrightarrow \hat Y_{r}
\]
Note how the $\hat Z_{p}$ puts a hole in the $\leftrightarrow$ string if $q < p
< r$.

Finally, let us consider the case with no contraction, $p \neq q \neq r \neq s$,
\begin{equation}
H^{\alpha \alpha}
\leftarrow
\sum_{p < q < r < s}
2
\langle pq || rs \rangle 
p^+ q^+ s r
\end{equation}
\[
+
2
\langle pr || qs \rangle
p^+ r^+ s q
+
2
\langle ps || qr \rangle
p^+ s^+ r q
\]
This can be obtained by enumerating all 24 permutations of $pqrs$ and then
simplifying to 3 terms by the 8-fold antisymmetry of the integrals/composition
operators, or by enumerating all 24 permutations of $pqrs$ and sieving by $p' <
q'$ and $r' < q'$ restrictions in the bra/ket of the antisymmetrized integrals.
Expanding the antisymmetrized integrals,
\begin{equation}
\hat H^{\alpha \alpha} 
\leftarrow
2
\sum_{p < q < r < q}
\left [
(pr|qs)
-
(ps|qr)
\right ]
p^{+}
q^{+}
s
r
\end{equation}
\[
+
\left [
(pq|rs)
-
(ps|rq)
\right ]
p^{+}
r^{+}
s
q
+
\left [
(pq|sr)
-
(pr|qs)
\right ]
p^{+}
s^{+}
r
q
\]
\[
2 
\sum_{p < q < r < s}
(pq|rs) 
[
p^+ r^+ s q
+
p^+ s^+ r q
]
\]
\[
+
(pr|qs)
[
p^+ q^+ s r
-
p^+ s^+ r q
]
\]
\[
+
(ps|qr)
[
-
p^+ q^+ s r
-
p^+ r^+ s q
]
\]
\[
=
\frac{1}{4}
\sum_{p < q < r < q}
(pq|rs) 
[
X_{p} \leftrightarrow X_{q} X_{r} \leftrightarrow X_{s}
+
X_{p} \leftrightarrow X_{q} Y_{r} \leftrightarrow Y_{s}
\]
\[
+
Y_{p} \leftrightarrow Y_{q} X_{r} \leftrightarrow X_{s}
+
Y_{p} \leftrightarrow Y_{q} Y_{r} \leftrightarrow Y_{s}
]
\]
\[
+
(pr|qs)
[
-
X_{p} \leftrightarrow X_{q} Y_{r} \leftrightarrow Y_{s}
+
X_{p} \leftrightarrow Y_{q} Y_{r} \leftrightarrow X_{s}
\]
\[
+
Y_{p} \leftrightarrow X_{q} X_{r} \leftrightarrow Y_{s}
-
Y_{p} \leftrightarrow Y_{q} X_{r} \leftrightarrow X_{s}
]
\]
\[
+
(ps|qr)
[
-
X_{p} \leftrightarrow X_{q} X_{r} \leftrightarrow X_{s}
-
X_{p} \leftrightarrow Y_{q} Y_{r} \leftrightarrow X_{s}
\]
\[
-
Y_{p} \leftrightarrow X_{q} X_{r} \leftrightarrow Y_{s}
-
Y_{p} \leftrightarrow Y_{q} Y_{r} \leftrightarrow Y_{s}
]
\]
\[
=
\frac{1}{4}
% 1
\sum_{p < q < r < s}
[
(pq|rs) 
- 
(ps|qr)
]
X_{p} \leftrightarrow X_{q} X_{r} \leftrightarrow X_{s}
\]
% 2
\[
+
[
(pq|rs) 
- 
(pr|qs)
]
X_{p} \leftrightarrow X_{q} Y_{r} \leftrightarrow Y_{s}
\]
% 5
\[
+
[
(pr|qs) 
- 
(ps|qr)
]
X_{p} \leftrightarrow Y_{q} Y_{r} \leftrightarrow X_{s}
\]
% 6
\[
+
[
(pr|qs) 
- 
(ps|qr)
]
Y_{p} \leftrightarrow X_{q} X_{r} \leftrightarrow Y_{s}
\]
% 3
\[
+
[
(pq|rs) 
- 
(pr|qs)
]
Y_{p} \leftrightarrow Y_{q} X_{r} \leftrightarrow X_{s}
\]
% 4
\[
+
[
(pq|rs) 
- 
(ps|qr)
]
Y_{p} \leftrightarrow Y_{q} Y_{r} \leftrightarrow Y_{s}
\]

Aggregrating,
\begin{equation}
\hat H^{\alpha \alpha}
=
\end{equation}
\[
\frac{1}{4}
\sum_{p < q}
\langle pq || pq \rangle
\]
\[
-
\frac{1}{4}
\sum_{p}
\left [
\sum_{q}
\langle pq || pq \rangle
\right ]
\hat Z_{p}
\]
\[
+
\frac{1}{4}
\sum_{p < q}
\langle pq || pq \rangle
\hat Z_{p} 
\otimes 
\hat Z_{q}
\]
\[
\frac{1}{4}
\sum_{q < r}
\left [
\sum_{p}
\langle pq || pr \rangle
\right ]
\hat X_{q} \leftrightarrow \hat X_{r}
\]
\[
+
\frac{1}{4}
\sum_{q < r}
\left [
\sum_{p}
\langle pq || pr \rangle
\right ]
\hat Y_{q} \leftrightarrow \hat Y_{r}
\]
\[
-
\frac{1}{4}
\sum_{q < r}
\sum_{p \neq q, r}
\langle pq || pr \rangle
\hat Z_{p}
\hat X_{q} \leftrightarrow \hat X_{r}
\]
\[
-
\frac{1}{4}
\sum_{q < r}
\sum_{p \neq q, r}
\langle pq || pr \rangle
\hat Z_{p}
\hat Y_{q} \leftrightarrow \hat Y_{r}
\]
\[
% 1
+
\frac{1}{4}
\sum_{p < q < r < s}
[
(pq|rs) 
- 
(ps|qr)
]
X_{p} \leftrightarrow X_{q} X_{r} \leftrightarrow X_{s}
\]
% 2
\[
+
\frac{1}{4}
\sum_{p < q < r < s}
[
(pq|rs) 
- 
(pr|qs)
]
X_{p} \leftrightarrow X_{q} Y_{r} \leftrightarrow Y_{s}
\]
% 5
\[
+
\frac{1}{4}
\sum_{p < q < r < s}
[
(pr|qs) 
- 
(ps|qr)
]
X_{p} \leftrightarrow Y_{q} Y_{r} \leftrightarrow X_{s}
\]
% 6
\[
+
\frac{1}{4}
\sum_{p < q < r < s}
[
(pr|qs) 
- 
(ps|qr)
]
Y_{p} \leftrightarrow X_{q} X_{r} \leftrightarrow Y_{s}
\]
% 3
\[
+
\frac{1}{4}
\sum_{p < q < r < s}
[
(pq|rs) 
- 
(pr|qs)
]
Y_{p} \leftrightarrow Y_{q} X_{r} \leftrightarrow X_{s}
\]
% 4
\[
+
\frac{1}{4}
\sum_{p < q < r < s}
[
(pq|rs) 
- 
(ps|qr)
]
Y_{p} \leftrightarrow Y_{q} Y_{r} \leftrightarrow Y_{s}
\]

\end{document}